\documentclass[prl, twocolumn, superscriptaddress, eprintnumbers, amsmath,amssymb]{revtex4-2}
\usepackage[]{graphicx}
\usepackage{epstopdf}
\usepackage{bm}
\usepackage{framed}

\usepackage{xcolor} 
\usepackage{dcolumn}

\begin{document}
\preprint{APS/123-QED}
\title{Domain-induced control of latent heat in freestanding BaTiO$_3$ membranes}

\author{Tapas Bar} \email{tapas.bar@icn2.cat}
\affiliation{Catalan Institute of Nanoscience and Nanotechnology (ICN2), CSIC and BIST,
 Campus UAB, 08193 Bellaterra, Barcelona, Spain}

\author{David Pesquera}
\affiliation{Catalan Institute of Nanoscience and Nanotechnology (ICN2), CSIC and BIST,
 Campus UAB, 08193 Bellaterra, Barcelona, Spain}
 
\author{Arnau Villalobos-Martin}
\affiliation{Catalan Institute of Nanoscience and Nanotechnology (ICN2), CSIC and BIST,
 Campus UAB, 08193 Bellaterra, Barcelona, Spain}
\affiliation{Department of Physics, Facultat de Ciencies, Universitat Autonoma de Barcelona,
 08193 Bellaterra, Barcelona, Spain}

\author{Cristian Rodriguez-Tinoco}
\affiliation{Catalan Institute of Nanoscience and Nanotechnology (ICN2), CSIC and BIST,
 Campus UAB, 08193 Bellaterra, Barcelona, Spain}
\affiliation{Department of Physics, Facultat de Ciencies, Universitat Autonoma de Barcelona,
 08193 Bellaterra, Barcelona, Spain}

\author{Umair Saeed}
\affiliation{Catalan Institute of Nanoscience and Nanotechnology (ICN2), CSIC and BIST,
 Campus UAB, 08193 Bellaterra, Barcelona, Spain}

\author{Kumara Cordero-Edwards}
\affiliation{Catalan Institute of Nanoscience and Nanotechnology (ICN2), CSIC and BIST,
 Campus UAB, 08193 Bellaterra, Barcelona, Spain}

\author{Jessica Padilla}
\affiliation{Catalan Institute of Nanoscience and Nanotechnology (ICN2), CSIC and BIST,
 Campus UAB, 08193 Bellaterra, Barcelona, Spain}

\author{José Manuel Caicedo Roque}
\affiliation{Catalan Institute of Nanoscience and Nanotechnology (ICN2), CSIC and BIST,
 Campus UAB, 08193 Bellaterra, Barcelona, Spain}

\author{José Santiso}
\affiliation{Catalan Institute of Nanoscience and Nanotechnology (ICN2), CSIC and BIST,
 Campus UAB, 08193 Bellaterra, Barcelona, Spain}

\author{Pol Lloveras}
\affiliation{Grup de Caracterizació de Materials, Departament de Física, EEBE and Barcelona Research Center in Multiscale Science and Engineering, Universitat Politècnica de Catalunya, Eduard Maristany, 10-14, 08019 Barcelona, Catalonia, Spain}%

\author{Léo Boron}
\affiliation{Laboratory of Condensed Matter Physics, University of Picardie, Jules Verne, 33 rue Saint Leu, Amiens, 80039, France}%

\author{Igor Lukyanchuk}\email{lukyanc@ferroix.net}
\affiliation{Laboratory of Condensed Matter Physics, University of Picardie, Jules Verne, 33 rue Saint Leu, Amiens, 80039, France}%

\author{Gustau Catalan} \email{gustau.catalan@icn2.cat}
\affiliation{Catalan Institute of Nanoscience and Nanotechnology (ICN2), CSIC and BIST,
 Campus UAB, 08193 Bellaterra, Barcelona, Spain}

\author{Javier Rodriguez-Viejo} \email{javier.rodriguez@icn2.cat}
\affiliation{Catalan Institute of Nanoscience and Nanotechnology (ICN2), CSIC and BIST,
 Campus UAB, 08193 Bellaterra, Barcelona, Spain}
\affiliation{Department of Physics, Facultat de Ciencies, Universitat Autonoma de Barcelona,
 08193 Bellaterra, Barcelona, Spain}

\begin{abstract}
Thin ferroelectric BaTiO$_3$ films often exhibit continuous transitions instead of the first-order behavior of bulk crystals, a discrepancy usually attributed to epitaxial strain or dimensionality. Using quasi-adiabatic nanocalorimetry on freestanding BaTiO$_3$ membranes—free of clamping and substrate heat sinking—we show that domain morphology, not thickness or boundary conditions, controls the transition order. Thick membranes with large, monodomain-like regions display clear latent heat, whereas thinner membranes with dense 180° domain patterns show a continuous transition despite undergoing the same tetragonal–cubic structural change confirmed by x-ray diffraction. Piezoresponse force microscopy links this behavior to domain-size evolution, and a Ginzburg–Landau analysis demonstrates how reduced domain size lowers the free-energy barrier, rounding a nominally first-order instability. These results identify domain morphology as the key determinant of ferroelectric transition order in oxide membranes and establish design guidelines for enhancing caloric effects through domain engineering.

\textbf{Summary Statement: In freestanding ferroelectric films, the order of the phase transition is not determined by strain or thickness, but by domains.}

\end{abstract}

\maketitle
Ferroelectric oxides have captivated the attention of condensed matter physicists for over a century, yielding an increasing variety of applications that include data storage, capacitors, electromechanical transducers and electrooptic devices \cite{Xu_Book13, Scott_Book18}. In ferroelectrics, the transition from the polar ferroelectric phase to the paraelectric phase is accompanied by a structural symmetry change which, in bulk samples, is first-order, with latent heat and thermal hysteresis \cite{Xu_Book13, Strukov_Book12}. This latent heat is an important property as it provides the majority of the cooling power in electrocaloric applications \cite{Mathur_AdvMat13, ValesCastro_prb21}, yet this latent heat exists precariously in archetypal perovskite ferroelectrics such as BaTiO$_3$, which are only weakly first-order and whose transition has in fact long been controversial  \cite{Senn_prl16, Stern_prl04, Zalar_prl03, Qi_prb16, Pasciak_prl18, Luo_prb23, Pramanick_prb15, Zhong_prl94}. Contributing to the complexity of the problem is that, in epitaxial thin films, the ferroelectric transition becomes second order \cite{Pertsev_prl98, Choi_Science04, Strukov_FerroE07, Scott_JAP08}. 

The change in the order of the ferroelectric phase transition in thin film ferroelectrics has been attributed to epitaxial clamping \cite{Pertsev_prl98, Choi_Science04, Strukov_FerroE07}, enhanced surface-to-volume ratio \cite{Parker_apl02, Binder_prb79, Sun_prl23}, strain gradients\cite{Catalan_Jpcm04, Yudin_prr21}, disorder, and low dimensionality  \cite{Imry_prb79, Aizenman_prl89, Cardy_prl97} . On the other hand, the change from first to second order is NOT inherent in ferroelectric thin films, as shown by an important counter-factual observation: thin lamellae ($<$100 nm thick), released from bulk single crystals using focused ion beam (FIB), yield sharp first-order transitions \cite{Saad_jpcm04}. Therefore, thickness alone does not necessarily suppress the latent heat, at least down to 100 nm. 

This unresolved state of affairs calls for separating intrinsic size effects from other concomitant contributions (clamping stress, strain relaxation gradients, etc) in the ferroelectric phase transition of thin films. Separating intrinsic size effects from substrate-induced effects in oxide thin films is facilitated the recent development of oxide membranes released via epitaxial lift-off  \cite{Lu_NatMat16, David_AdvMat20, Pesquera_JPCM22}. Yet, while the functional properties of such membranes have been  characterized \cite{Ganguly_AdvMat24, Lee_AdvMat22, Sun_prl23}, there are no direct measurements of their latent heat. In this article,  we use nano-calorimetry \cite{Allen_ThermoActa04, Rodríguez-Viejo_Book16, Vila-Costa_prl20} to characterize free-standing BaTiO$_3$ membranes, addressing the origin (and the disappearance) of latent heat in ferroelectric phase transitions. We find that size can affect the order of the transition, but indirectly: what happens is that thin films develop antiparallel domains to minimize depolarization, and it is the appearance of 180 degree domain walls that "kills" the latent heat. Besides the experimental findings, we will provide a Landau-type thermodynamic formalism that explains why and how domains rule the order of the transition.

\begin{figure}
\centering
\includegraphics[width=0.45\textwidth]{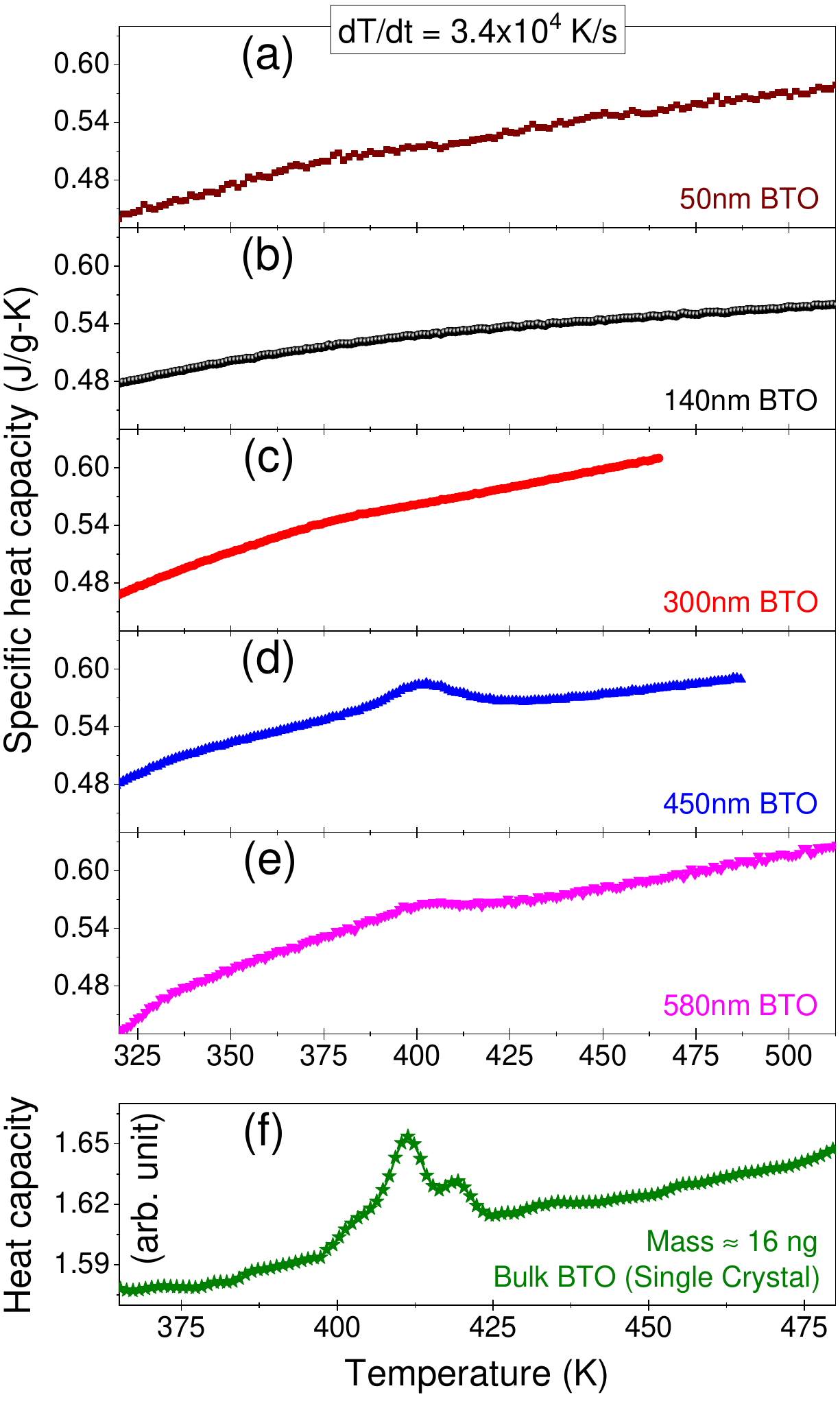}
\caption[0.5\textwidth]{Temperature-dependent specific heat capacity of BaTiO$_3$ membranes of different thicknesses, obtained from nanocalorimetry measurements (a, b, c, d, e). The data for a bulk crystal, measured in the same setup, is shown in (f) in arbitrary units. The heating rates for the data shown are approximately 3.4 $\times 10^{4}$ K/s. Measurement details are provided in the supplementary material.}
\label{fig:fig1}
\end{figure}

Epitaxial  films of BaTiO$_3$ with out-of-plane (001) orientation were grown on GdScO$_3$ substrates with a 7-nm-thick sacrificial buffer layer of Sr$_3$Al$_2$O$_6$ (SAO) that was then dissolved in water. The samples were grown using pulsed laser deposition at 750$^\circ$C under 100 mTorr oxygen pressure. The released membranes were then transferred to a nano-calorimetric chip using a polymer-based transfer method \cite{Pesquera_JPCM22}.  The calorimetric chips consist of a 150 nm platinum-based heater (also functioning as a thermometer) deposited on a 450 nm free-standing silicon nitride (Si$_3$N$_4$) membrane supported by a Si frame. Further details of the sample preparation and transfer procedures are provided in the supplementary materials \cite{supplementary}. In high vacuum, a small electrical current (20–56 mA) provided to the chip rapidly increased the temperature of the sensing area at rates of $\sim$0.1–5 $\times$10$^{4}$ K/s. Under such conditions, heat losses are negligible, and the measured heat capacity can be regarded as nearly adiabatic. The heat capacity of the sample was obtained by subtracting the baseline heat capacity of the calorimetric chip. The true adiabatic heat capacity can also be extracted by correcting for finite-rate heat losses, determined through linear extrapolation of rate-dependent measurements, as presented in the supplementary materials. A brief overview of the instrumentation is provided there \cite{supplementary}, with detailed descriptions available in earlier publications \cite{Allen_ThermoActa04, Rodríguez-Viejo_Book16, Vila-Costa_prl20}. The calorimetry results of our BaTiO$_3$  membranes measured using this method is shown in Fig. \ref{fig:fig1}.

The results show that samples thinner than 400 nm have no specific heat anomaly [Fig. \ref{fig:fig1} (a-c)], while thicker membranes do display latent heat peaks [Fig. \ref{fig:fig1}(d-e)]. For the 450-nm membrane, the latent heat is comparable to that of bulk single crystals, while the 570-nm membrane shows a smaller value (~40$\%$ of the bulk). As a reference for comparison, we also measured a small  piece (mass $\approx 16$ ng, volume $\approx 19\times14\times10\ \mu m^3$) of cystal BaTiO$_3$, cut from a larger piece using FIB and measured at the same heating rate as for membranes. Since the tungsten alloy used during FIB process introduces an extra contribution to the heat capacity, we chose to plot the heat capacity in an arbitrary unit [Fig. \ref{fig:fig1}(f)]. We point out that the mass of the FIB-cut single crystal was only 16.5 ng, which is much smaller than the mass of the thinnest membrane in our study (650 ng for the 140-nm-thick membrane), and therefore the lack of visible latent heat in the measurements is not due to lack of instrumental resolution. The results thus indicate that the thicker membranes have a first order phase transition (with latent heat), and the thinner membranes have either second order or no transition at all. 

\begin{figure}
\centering
\includegraphics[width=0.48\textwidth]{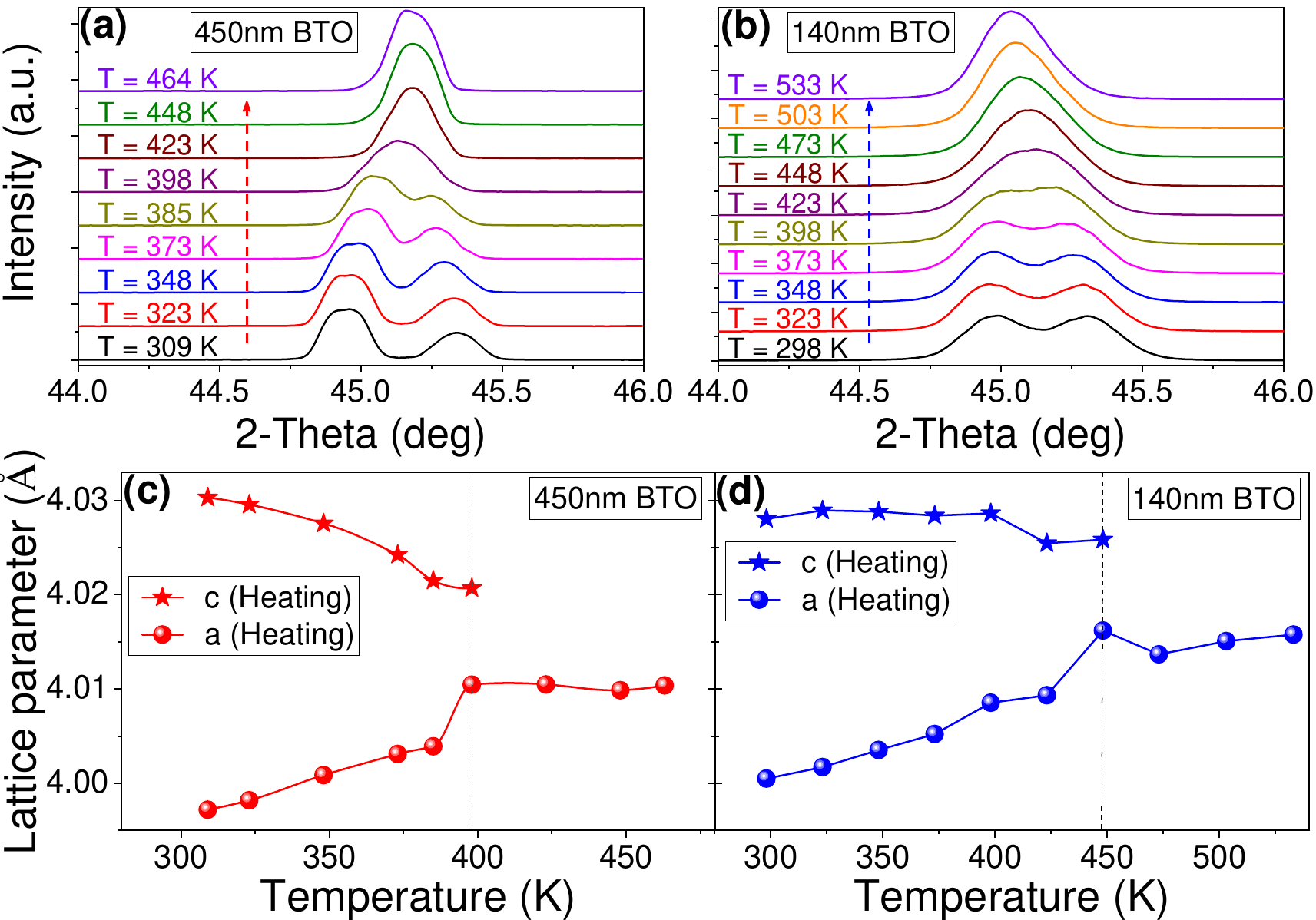}
\caption[0.5\textwidth]{Temperature-dependent X-ray diffraction patterns of 450 nm BTO (a) and 140 nm BTO (b) membranes placed on nanocalorimeter chips. (c) The lattice parameters of the membranes were extracted from $\theta$–2$\theta$ scans around the (002) and (200) diffraction peaks.
}
\label{fig:XRD}
\end{figure}

To determine whether the thinner samples (those not displaying latent heat peak) still have a phase transition at all, we conducted X-ray diffraction measurements.  Starting at room temperature, the samples were heated at a rate of 5 K/min and allowed to stabilize for 20 minutes at each temperature before measuring $\theta-2\theta$ scans around the (002) and (200) diffraction peaks, corresponding to the $c-$oriented and $a-$oriented tetragonal domains characteristic of the ferroelectric phase [Fig. \ref{fig:XRD} (a, b)]. The \textit{a} and \textit{c} lattice parameters were then calculated by fitting (200) and (002) diffraction peaks respectively, and plotted as a function of temperature [Fig. \ref{fig:XRD}(c, d)]. Upon increasing temperature, the two diffraction peaks collapse into one, signaling the structural transition to the paralectric phase. This structural transition was observed in all the membranes, irrespective of whether they had latent heat. Altogether, then, the experimental results indicate that all membranes have a ferroelectric transition, with a size-induced change from first to second order that is not caused by epitaxial strain.

 The results thus indicate a size effect. However, this effect is unlikely to be intrinsic, because our thinnest membrane (140 nm) is still thicker than the single-crystal lamellae where Saad et al observed a first-order transition \cite{Saad_jpcm04}. The membranes are also too thick to be considered 2-D, so dimensional crossover cannot be invoked and, being clamping-free, nor can we invoke epitaxial strain. The relationship between phase transition order and film thickness therefore eludes existing models, and a new explanation is required. 

\begin{figure}
\centering
\includegraphics[width=0.49\textwidth]{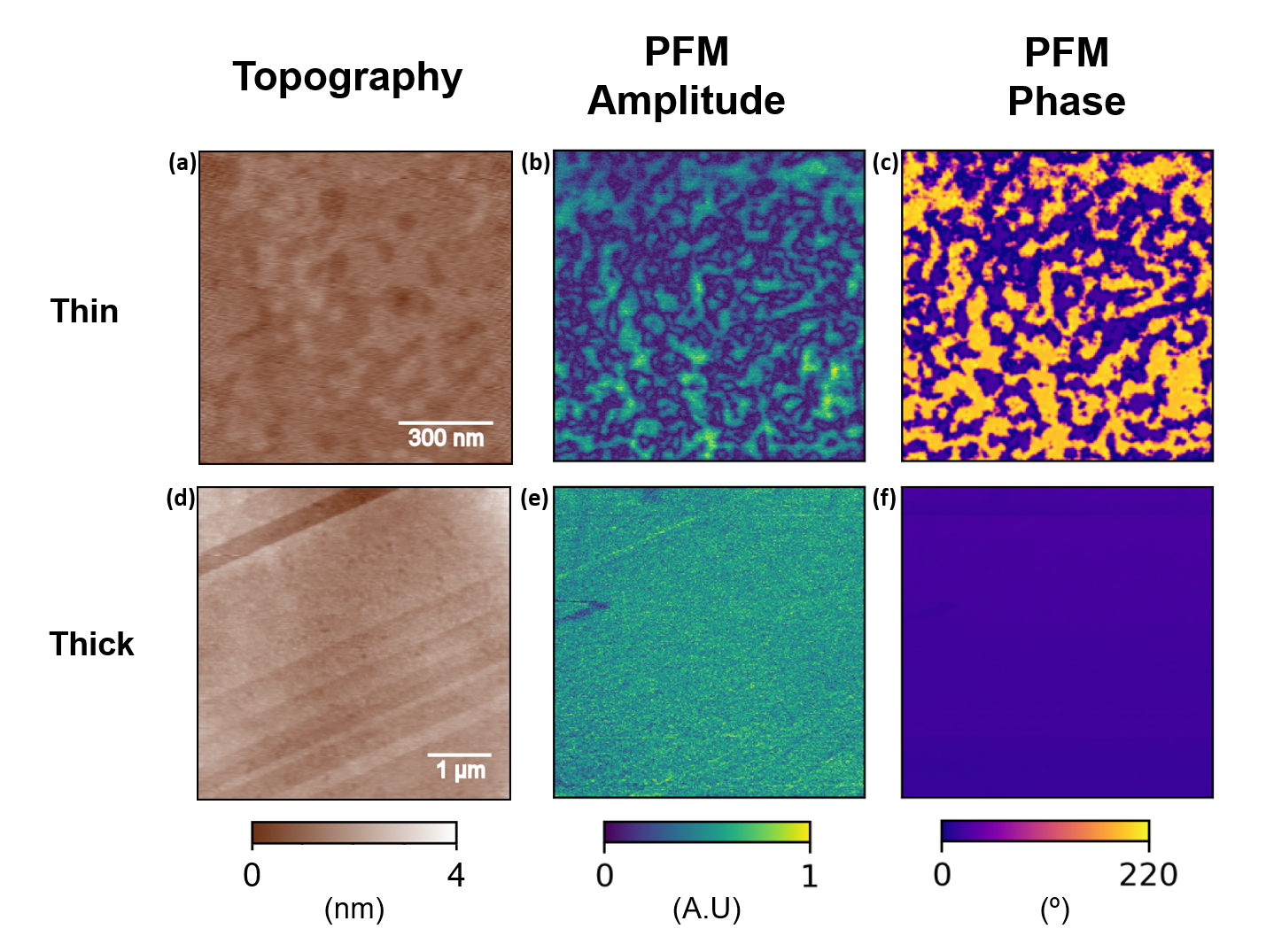}
\caption[0.5\textwidth]{Piezoresponse of BaTiO$_3$ (BTO) membranes for a thin sample (without a latent heat peak) and a thick sample (showing latent heat). All scans were performed at room temperature.}
\label{fig:PFM}
\end{figure}

To gain further insight, we have  examined the films' ferroelectric domain configuration using piezoresponse force microscopy (PFM), which provides information on the magnitude and orientation of the polarization through its amplitude and phase signals, respectively. Vertical PFM (VPFM) was performed at room temperature in dual AC resonance tracking (DART) mode \cite{Rodriguez_NanoTech2007} using an AC excitation voltage of 700 mV. The measurements mapped the out-of-plane polarization of two representative membranes: one without detectable latent heat (300 nm) and another showing a clear latent heat peak (450 nm). Figures \ref{fig:PFM}(a–c) display, respectively, the topography, VPFM amplitude, and phase of the thinner membrane, and Figs. \ref{fig:PFM}(d–f) correspond to the thicker one. The thinner membrane exhibits a dense arrangement of labyrinthine domains with 180° phase contrast and amplitude minima at the domain boundaries —features characteristic of oppositely-polarized ferroelectric regions. In contrast, the thicker membrane exhibits uniform amplitude and phase, consistent with a homogeneous polarization state (the topography also reveals some straight-line features along the $<100>$ direction, possibly due to sparse \textit{a}-\textit{c} twins). The results thus suggest to a correlation between domain morphology and order of the phase transition: small 180$^{\circ}$ domains are correlated with a second-order transition, while homogeneous polarization is correlated with a first-order transition. To investigate whether this correlation is coincidental or has a physical basis, we employ a phenomenological Ginzburg–Landau model.

The starting point for the description of any ferroelectric phase transition is a polynomial expansion of the free energy as a function of polarization: 
In the Ginzburg--Landau approach, the thermodynamics of the transition in ferroelectrics can be expressed in terms of a polarization order parameter $\mathbf{P}$, which in the simplest case reduces to its scalar amplitude $P$, as in a uniaxial ferroelectric. The corresponding free energy is then
\begin{equation}
F=\int dV\left[ A P^2 + B P^4 + C P^6 \right],
\label{BasicF}
\end{equation}
where $A=\alpha(T-T_c)$ changes sign at the Curie temperature $T_c$, and the higher-order coefficients determine the order of the transition and $C>0$. For $B<0$, the transition is first order, with a discontinuous jump of the order parameter and latent heat
$L \simeq T_c\,\alpha|B|/2C$.
For $B>0$, the transition is continuous and is characterized instead by a heat-capacity jump
$\Delta C=T_c\alpha^2/2B$.
Thus, within this effective description, the distinction between first- and second-order behavior is governed by the sign of the quartic coefficient. As shown below, this coefficient is renormalized by the domain configuration.

For cubic BaTiO$_3$, this effective scalar description follows from the full Ginzburg--Landau--Devonshire functional for the vector polarization $\mathbf{P}$ coupled to elasticity. The polarization part is
\begin{equation}
F_{\mathrm P}
=
\int dV
\left[
a_i(T) P_i^2
+
a_{ij} P_i^2 P_j^2
+
a_{ijk} P_i^2 P_j^2 P_k^2
\right]_{i \le j \le k},
\label{FP}
\end{equation}
and the elastic part is
\begin{equation}
F_{\mathrm{elast}}
=
\int dV
\left[
- Q_{ijkl} P_i P_j \sigma_{kl}
-\frac{1}{2} S_{ijkl}\sigma_{ij}\sigma_{kl}
\right].
\label{Felast}
\end{equation}
Here $Q_{ijkl}$ is the electrostrictive tensor and $S_{ijkl}$ the elastic compliance tensor.
Standard material parameters for BaTiO$_3$ are taken from Ref.~\cite{Book_Rabe07} (See also End Matter).

In the \textit{monodomain state}, with polarization oriented along $z$, $\mathbf{P}=(0,0,P)$, and in the absence of external stress, the elastic contribution vanishes and Eqs.~(\ref{FP}) and (\ref{Felast}) reduce to the scalar form of Eq.~(\ref{BasicF}), with $A=a_1$, $B=a_{11}$, and $C=a_{111}$.  
Importantly, because the coefficient $a_{11}$ in BaTiO$_3$ is negative near the transition, the monodomain transition is first order.

In the Landau-Kittel \textit{multidomain state}, induced by the depolarization forces, domain walls close to the transition broaden~\cite{Chrosch_jap99,Shih_prb94} and span the full domain period, so that a dense 180$^\circ$ domain pattern is naturally described by a smooth sinusoidal polarization texture~\cite{Stephanovich_Ferroelectics03, Stephanovich_prl05, Lukyanchuk_prl09}, rather than by sharp interfaces. More specifically, depolarization-induced periodic domains~\cite{Bratkovsky_prl00, Zubko_prl10} can be viewed as alternating clockwise and counterclockwise polarization vortices in the $(x,z)$ plane, extended along $y$~\cite{Kornev_prl04,Yadav_Nature16}. A convenient representation of such a state is~\cite{Lukyanchuk_arXiv24}
\begin{gather}
\mathbf{P}(\mathbf{r})=\gamma P\sin\left( \kappa_x x \right) \sin \left(\kappa_z z\right)\mathbf{e}_{x}
+ P\cos\left(\kappa_x x \right) \cos \left(\kappa_z z\right)\mathbf{e}_{z},
\notag \\
\kappa_x=2\pi / d, \qquad \kappa_z=\pi / h, \qquad \gamma=\kappa_z / \kappa_x= d / 2h,
\label{SoftDomain}
\end{gather}
commonly referred to as a soft-domain profile~\cite{Lukyanchuk_prl09}. Here $h$ denotes the film thickness and $d$ the domain half-period. This form provides a smooth polarization texture, enforces tangential alignment at the film surfaces ($z=\pm h/2$), and satisfies $\nabla\!\cdot\!\mathbf{P}=0$, thereby eliminating bulk bound charge while avoiding surface bound charge. The gradient-energy contribution is neglected here, since it is smaller than the local Landau terms by a factor of order $(\xi/d)^2\sim 10^{-3}$--$10^{-4}$ with $\xi\sim1$--$2$~nm and $d\sim50$--$100$~nm.

In the multidomain state, both the quadratic and quartic coefficients are renormalized, with separate contributions from polarization inhomogeneity and elasticity:
\begin{gather}
A=\Bar A_{\mathrm {P}}+\Bar A_{\mathrm {elast}}, 
\quad
B=\Bar B_{\mathrm {P}}+\Bar B_{\mathrm {elast}}.
\label{AB}
\end{gather}
The detailed derivation is given in the End Notes; the code for their calculation for arbitrary material parameters is provided in~\cite{LukyanchukBoron2026Zenodo}.
Here we summarize the resulting expressions.

For the quadratic term, one obtains 
\begin{gather}
\Bar A_{\mathrm {P}}=\frac{1}{4}a_1(1+\gamma^2),
\label{AP} \\
\Bar A_{\mathrm {elast}}=-\frac{1}{4}\,2Q_{1122}\,\sigma_s
\label{Ael}
\end{gather}
where $\Bar A_{\mathrm P}$ results from averaging the multidomain polarization texture, while $\Bar A_{\mathrm {elast}}$ arises from the macroscopic in-plane stress $\sigma_s$=$\sigma_{xx}$=$\sigma_{yy}$~\cite{Kondovych2025}. Compressive in-plane stress shifts the transition temperature upward. Such stress may, for instance, originate from surface tension, for which $\sigma_s=2\tau/h$, with $\tau\simeq1$--$50\,\mathrm{N\,m^{-1}}$~\cite{Uchino1989,Cammarata1994,Ma2009,Diehm2012,Kondovych2025} the effective surface-tension coefficient, or from a degraded, processing-induced near-surface layer~\cite{Lukyanchuk_prb09}. In these cases, $\sigma_s$ scales as $1/h$ and is therefore significant mostly in sufficiently thin membranes.

For the quartic term, one obtains 
\begin{gather}
\Bar B_{\mathrm {P}}=\frac{1}{64}\left[9a_{11}(1+\gamma^4)+a_{12}\gamma^2\right],
\label{BP} \\
\Bar B_{\mathrm {elast}}
\approx(4.2+2.4\gamma^2+4.3\gamma^4)\times 10^7\,\text{C}^{-4}\text{m}^6\text{N}.
\label{Bel}
\end{gather} 

The polarization contribution $\Bar B_{\mathrm P}$ results from averaging the quartic polarization energy over the multidomain texture. In the monodomain state, this term is controlled solely by $a_{11}<0$. In the multidomain vortex-like state, however, the polarization acquires both in-plane and out-of-plane components, so that the quartic energy contains, in addition to the negative $a_{11}$ term, a mixed term proportional to 
the positive $a_{12}=3.2\times 10^8$\,C$^{-4}$m$^6$N.

Meanwhile, the elastic term $\Bar B_{\mathrm{elast}}$ provides a further positive contribution through electrostriction, whereby the nonuniform polarization texture generates a distributed internal source of elastic stress. Complete local stress compensation is impossible because the stress field must satisfy mechanical equilibrium, i.e., local force balance. The elastic response is therefore nonlocal and cannot relax this source point by point, leaving a nonzero total internal stress whose positive self-energy is proportional to $P^4$.

The calculated quartic coefficient $B$ is plotted in Fig.~4 as a function of $\gamma$ for temperatures near the transition temperature. For the experimentally relevant range of $\gamma$, the coefficient changes sign at approximately $T=370$~K.

An intuitive way to understand these results is that the soft-domain texture is not a uniform out-of-plane polarization. Near the 180° walls, the polarization is rotated, so that there are mixed in-plane and out-of-plane components, such as $a_{12}P_x^2P_z^2$, which would be absent in the monodomain state with only
 $P_z$. Additionally, the decreased  out-of-plane polarization at the walls introduces, via electrostrictive coupling, a mechanical stress that has long-range consequences. This elastic energy contribution is positive and proportional to $P^4$, which pushes the effective coefficient $B_P$  toward positive values. Moreover,  as the phase transition is approached from below, the domain walls continously broaden  \cite{Chrosch_jap99, Shih_prb94}, thus facilitating a smooth transition to the paraelectric state.

With these renormalized coefficients, and with the sixth-order term, as shown in the End Notes, modified only by the polarization inhomogeneity,
\begin{equation}
\Bar C=\frac{1}{256}\left(25a_{111}(1+\gamma^6)+a_{112}\gamma^2(1+\gamma^2)\right),
\label{C}
\end{equation}

the thermodynamic behavior expected for thick and thin membranes can now be compared quantitatively, as summarized in Table~\ref{TableCoeff}. For the thicker, monodomain membranes, the effective quartic coefficient $B$ is negative, yielding a first-order transition and finite latent heat. For the thinner, polydomain membranes, however, it becomes positive, yielding a continuous transition with vanishing latent heat, as observed.

\begin{table}[t]
\centering
\begin{ruledtabular}
\begin{tabular}{lcc}
 & {Thick films} & {Thin film} \\
     & $h=$200--500~nm, & $h=140$~nm,   \\
      & monodomain state, & multidomain state,   \\
  & 1st order transition& 2nd order transition  \\
\hline
\\[-0.6em]
 $T'_c$  
& $390$--$400$~K 
& $\approx 430$~K \\

$\alpha$ (C$^{-2}$m$^{2}$NK$^{-1}$) 
& $3.34\times 10^5$ 
& $0.90\times 10^5$ \\

$B$ (C$^{-4}$m$^{6}$N) 
& $-1.69\times10^{8}$ 
& $0.40\times10^{8}$ \\

$C$ (C$^{-6}$m$^{10}$N) 
& $2.37\times10^{9}$ 
& $0.08\times10^{9}$ \\

 $L$ (J\,g$^{-1}$) 
& $0.75$ 
& $-$  \\

$ \Delta C$ (J\,g$^{-1}$K$^{-1}$) 
& $-$ 
& $7.2\times10^{-3}$ \\

\end{tabular}
\end{ruledtabular}
\caption{
Transition temperatures $T'_c$, effective Landau coefficients $\alpha$, $B$, and $C$, and the calculated caloric characteristics of the transition,  latent heat $L$, and the specific heat jump  $\Delta C$, for thick and thin free-standing BaTiO$_3$ films.
}
\label{TableCoeff}
\end{table}

\begin{figure}
    \centering
    \includegraphics[width=0.99\linewidth]{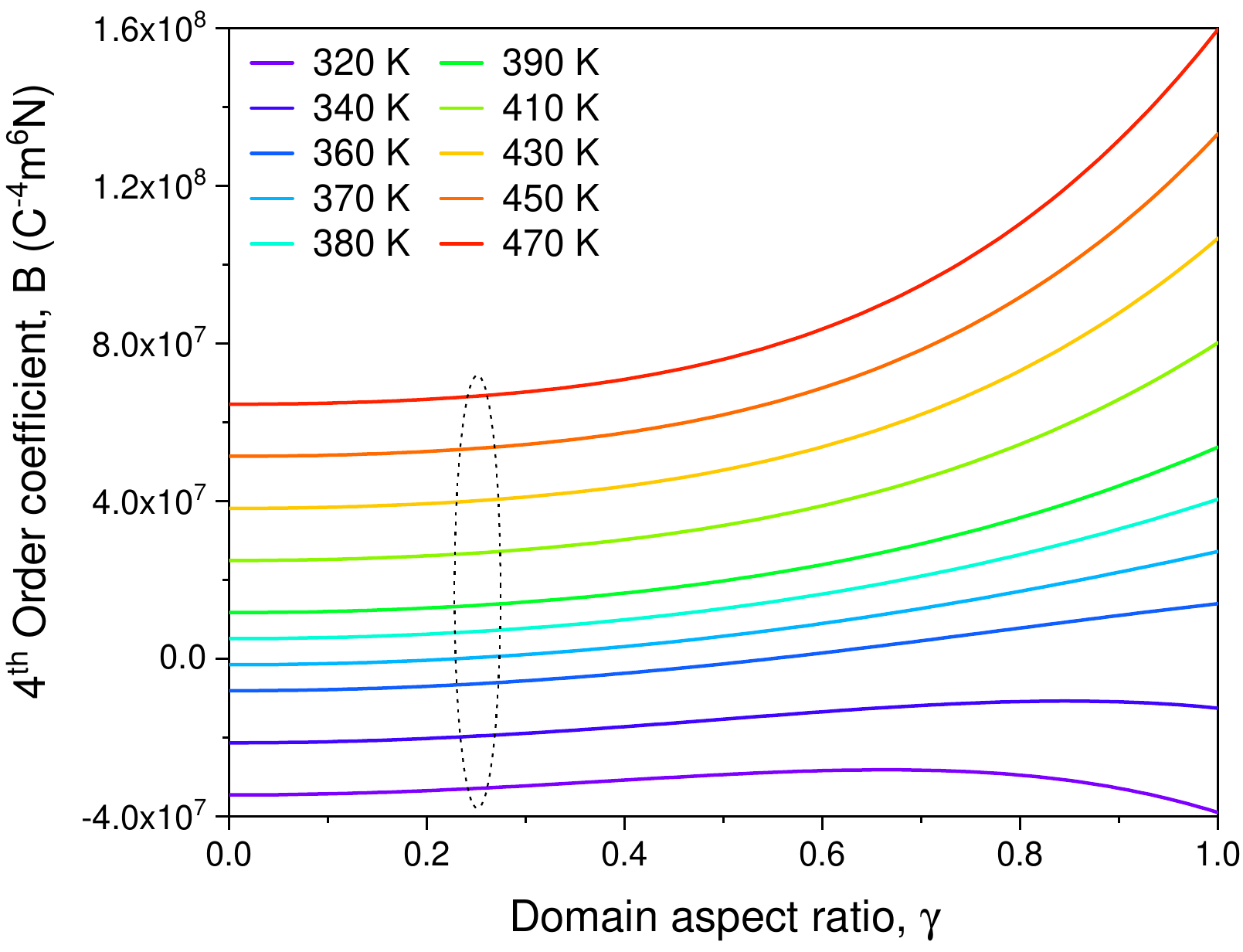}

  \caption[0.5\textwidth]{Fourth-order coefficient $B$ as a function of the domain aspect ratio $\gamma$ at different temperatures. The value $\gamma \approx 0.25$ for the 140~nm membrane is indicated by a dotted line.}
    \label{fig:fig4}
\end{figure}

In summary: thickness matters, but indirectly, via domain formation.  Thin freestanding membranes favor dense multidomain textures, whose polar and elastic renormalizations suppress the first-order character of the transition.  The good news for electrocaloric effects is that the loss of latent heat, which is inevitable in epitaxial films \cite{Pertsev_prl98, Catalan_Jpcm04}, is not inevitable in membranes: good screening of depolarizing fields via suitable electrodes and/or adsorbates can stabilize the monodomain state \cite{Saad_jpcm04, Mirzamohammadi_apl25} and thereby recover the first order phase transition and its concomitant latent heat, essential for giant electrocaloric effects.  Freestanding membranes thus offer a route to control the transition order and optimize electrocaloric response through domain engineering.

\section*{Acknowledgments}

We acknowledge Grant No. TED2021-131363B/I00 funded by MICIU/AEI/ 10.13039/501100011033 and by the European Union NextGenerationEU/PRTR. T.B. acknowledges Juan de la Cierva post-doctoral fellowship, Grant No. JDC2022-049886-I funded by MICIU/AEI/ 10.13039/501100011033 and by the European Union NextGenerationEU/PRTR. D.P. acknowledges the “Ramon y Cajal” Senior Research Fellowship (RYC-2023-044643-I) from the Spanish Ministry of Science, Innovation and Universities. D.P. and K.C.E. acknowledge grant PID2022-140589NB-I00, funded by MCIN/AEI/10.13039/501100011033. P.L. acknowledges Grant No. PID2023-146623NB-I00 funded by MICIU/AEI/ 10.13039/501100011033. G.C. acknowledges PID2023-148673NB-I00. I.L. and G.C. acknowledge the 101236483 — 3D-TOPO — HORIZON-MSCA-SE. The ICN2 is funded by the CERCA program/Generalitat de Catalunya. The ICN2 is supported by the Severo Ochoa program, grant  CEX2021-001214-S. T.B. J.R.-V acknowledge support from 2021SGR-00644 funded by AGAUR.

\bibliography{sample}

\begin{thebibliography}{56}%
\makeatletter
\providecommand \@ifxundefined [1]{%
 \@ifx{#1\undefined}
}%
\providecommand \@ifnum [1]{%
 \ifnum #1\expandafter \@firstoftwo
 \else \expandafter \@secondoftwo
 \fi
}%
\providecommand \@ifx [1]{%
 \ifx #1\expandafter \@firstoftwo
 \else \expandafter \@secondoftwo
 \fi
}%
\providecommand \natexlab [1]{#1}%
\providecommand \enquote  [1]{``#1''}%
\providecommand \bibnamefont  [1]{#1}%
\providecommand \bibfnamefont [1]{#1}%
\providecommand \citenamefont [1]{#1}%
\providecommand \href@noop [0]{\@secondoftwo}%
\providecommand \href [0]{\begingroup \@sanitize@url \@href}%
\providecommand \@href[1]{\@@startlink{#1}\@@href}%
\providecommand \@@href[1]{\endgroup#1\@@endlink}%
\providecommand \@sanitize@url [0]{\catcode `\\12\catcode `\$12\catcode
  `\&12\catcode `\#12\catcode `\^12\catcode `\_12\catcode `\%12\relax}%
\providecommand \@@startlink[1]{}%
\providecommand \@@endlink[0]{}%
\providecommand \url  [0]{\begingroup\@sanitize@url \@url }%
\providecommand \@url [1]{\endgroup\@href {#1}{\urlprefix }}%
\providecommand \urlprefix  [0]{URL }%
\providecommand \Eprint [0]{\href }%
\providecommand \doibase [0]{https://doi.org/}%
\providecommand \selectlanguage [0]{\@gobble}%
\providecommand \bibinfo  [0]{\@secondoftwo}%
\providecommand \bibfield  [0]{\@secondoftwo}%
\providecommand \translation [1]{[#1]}%
\providecommand \BibitemOpen [0]{}%
\providecommand \bibitemStop [0]{}%
\providecommand \bibitemNoStop [0]{.\EOS\space}%
\providecommand \EOS [0]{\spacefactor3000\relax}%
\providecommand \BibitemShut  [1]{\csname bibitem#1\endcsname}%
\let\auto@bib@innerbib\@empty
\bibitem [{\citenamefont {Xu}(2013)}]{Xu_Book13}%
  \BibitemOpen
  \bibfield  {author} {\bibinfo {author} {\bibfnamefont {Y.}~\bibnamefont
  {Xu}},\ }\href@noop {} {\emph {\bibinfo {title} {Ferroelectric materials and
  their applications}}}\ (\bibinfo  {publisher} {Elsevier},\ \bibinfo {year}
  {2013})\BibitemShut {NoStop}%
\bibitem [{\citenamefont {Huang}\ and\ \citenamefont
  {Scott}(2018)}]{Scott_Book18}%
  \BibitemOpen
  \bibfield  {author} {\bibinfo {author} {\bibfnamefont {H.}~\bibnamefont
  {Huang}}\ and\ \bibinfo {author} {\bibfnamefont {J.~F.}\ \bibnamefont
  {Scott}},\ }\href@noop {} {\emph {\bibinfo {title} {Ferroelectric materials
  for energy applications}}}\ (\bibinfo  {publisher} {John Wiley \& Sons},\
  \bibinfo {year} {2018})\BibitemShut {NoStop}%
\bibitem [{\citenamefont {Strukov}\ and\ \citenamefont
  {Levanyuk}(2012)}]{Strukov_Book12}%
  \BibitemOpen
  \bibfield  {author} {\bibinfo {author} {\bibfnamefont {B.~A.}\ \bibnamefont
  {Strukov}}\ and\ \bibinfo {author} {\bibfnamefont {A.~P.}\ \bibnamefont
  {Levanyuk}},\ }\href@noop {} {\emph {\bibinfo {title} {Ferroelectric
  phenomena in crystals: physical foundations}}}\ (\bibinfo  {publisher}
  {Springer Science \& Business Media},\ \bibinfo {year} {2012})\BibitemShut
  {NoStop}%
\bibitem [{\citenamefont {Moya}\ \emph {et~al.}(2013)\citenamefont {Moya},
  \citenamefont {Stern-Taulats}, \citenamefont {Crossley}, \citenamefont
  {González-Alonso}, \citenamefont {Kar-Narayan}, \citenamefont {Planes},
  \citenamefont {Mañosa},\ and\ \citenamefont {Mathur}}]{Mathur_AdvMat13}%
  \BibitemOpen
  \bibfield  {author} {\bibinfo {author} {\bibfnamefont {X.}~\bibnamefont
  {Moya}}, \bibinfo {author} {\bibfnamefont {E.}~\bibnamefont {Stern-Taulats}},
  \bibinfo {author} {\bibfnamefont {S.}~\bibnamefont {Crossley}}, \bibinfo
  {author} {\bibfnamefont {D.}~\bibnamefont {González-Alonso}}, \bibinfo
  {author} {\bibfnamefont {S.}~\bibnamefont {Kar-Narayan}}, \bibinfo {author}
  {\bibfnamefont {A.}~\bibnamefont {Planes}}, \bibinfo {author} {\bibfnamefont
  {L.}~\bibnamefont {Mañosa}},\ and\ \bibinfo {author} {\bibfnamefont {N.~D.}\
  \bibnamefont {Mathur}},\ }\bibfield  {title} {\bibinfo {title} {Giant
  electrocaloric strength in single-crystal $\mathrm{BaTiO}_{3}$},\ }\href
  {https://doi.org/https://doi.org/10.1002/adma.201203823} {\bibfield
  {journal} {\bibinfo  {journal} {Advanced Materials}\ }\textbf {\bibinfo
  {volume} {25}},\ \bibinfo {pages} {1360} (\bibinfo {year}
  {2013})}\BibitemShut {NoStop}%
\bibitem [{\citenamefont {Vales-Castro}\ \emph {et~al.}(2021)\citenamefont
  {Vales-Castro}, \citenamefont {Faye}, \citenamefont {Vellvehi}, \citenamefont
  {Nouchokgwe}, \citenamefont {Perpi{\~n}{\`a}}, \citenamefont {Caicedo},
  \citenamefont {Jord{\`a}}, \citenamefont {Roleder}, \citenamefont {Kajewski},
  \citenamefont {Perez-Tomas}, \citenamefont {Defay},\ and\ \citenamefont
  {Catalan}}]{ValesCastro_prb21}%
  \BibitemOpen
  \bibfield  {author} {\bibinfo {author} {\bibfnamefont {P.}~\bibnamefont
  {Vales-Castro}}, \bibinfo {author} {\bibfnamefont {R.}~\bibnamefont {Faye}},
  \bibinfo {author} {\bibfnamefont {M.}~\bibnamefont {Vellvehi}}, \bibinfo
  {author} {\bibfnamefont {Y.}~\bibnamefont {Nouchokgwe}}, \bibinfo {author}
  {\bibfnamefont {X.}~\bibnamefont {Perpi{\~n}{\`a}}}, \bibinfo {author}
  {\bibfnamefont {J.~M.}\ \bibnamefont {Caicedo}}, \bibinfo {author}
  {\bibfnamefont {X.}~\bibnamefont {Jord{\`a}}}, \bibinfo {author}
  {\bibfnamefont {K.}~\bibnamefont {Roleder}}, \bibinfo {author} {\bibfnamefont
  {D.}~\bibnamefont {Kajewski}}, \bibinfo {author} {\bibfnamefont
  {A.}~\bibnamefont {Perez-Tomas}}, \bibinfo {author} {\bibfnamefont
  {E.}~\bibnamefont {Defay}},\ and\ \bibinfo {author} {\bibfnamefont
  {G.}~\bibnamefont {Catalan}},\ }\bibfield  {title} {\bibinfo {title} {Origin
  of large negative electrocaloric effect in antiferroelectric
  $\mathrm{PbZr}{\mathrm{o}}_{3}$},\ }\href
  {https://doi.org/10.1103/PhysRevB.103.054112} {\bibfield  {journal} {\bibinfo
   {journal} {Phys. Rev. B}\ }\textbf {\bibinfo {volume} {103}},\ \bibinfo
  {pages} {054112} (\bibinfo {year} {2021})}\BibitemShut {NoStop}%
\bibitem [{\citenamefont {Senn}\ \emph {et~al.}(2016)\citenamefont {Senn},
  \citenamefont {Keen}, \citenamefont {Lucas}, \citenamefont {Hriljac},\ and\
  \citenamefont {Goodwin}}]{Senn_prl16}%
  \BibitemOpen
  \bibfield  {author} {\bibinfo {author} {\bibfnamefont {M.~S.}\ \bibnamefont
  {Senn}}, \bibinfo {author} {\bibfnamefont {D.~A.}\ \bibnamefont {Keen}},
  \bibinfo {author} {\bibfnamefont {T.~C.~A.}\ \bibnamefont {Lucas}}, \bibinfo
  {author} {\bibfnamefont {J.~A.}\ \bibnamefont {Hriljac}},\ and\ \bibinfo
  {author} {\bibfnamefont {A.~L.}\ \bibnamefont {Goodwin}},\ }\bibfield
  {title} {\bibinfo {title} {Emergence of long-range order in
  $\mathrm{BaTiO}_{3}$ from local symmetry-breaking distortions},\ }\href
  {https://doi.org/10.1103/PhysRevLett.116.207602} {\bibfield  {journal}
  {\bibinfo  {journal} {Phys. Rev. Lett.}\ }\textbf {\bibinfo {volume} {116}},\
  \bibinfo {pages} {207602} (\bibinfo {year} {2016})}\BibitemShut {NoStop}%
\bibitem [{\citenamefont {Stern}(2004)}]{Stern_prl04}%
  \BibitemOpen
  \bibfield  {author} {\bibinfo {author} {\bibfnamefont {E.~A.}\ \bibnamefont
  {Stern}},\ }\bibfield  {title} {\bibinfo {title} {Character of order-disorder
  and displacive components in barium titanate},\ }\href
  {https://doi.org/10.1103/PhysRevLett.93.037601} {\bibfield  {journal}
  {\bibinfo  {journal} {Phys. Rev. Lett.}\ }\textbf {\bibinfo {volume} {93}},\
  \bibinfo {pages} {037601} (\bibinfo {year} {2004})}\BibitemShut {NoStop}%
\bibitem [{\citenamefont {Zalar}\ \emph {et~al.}(2003)\citenamefont {Zalar},
  \citenamefont {Laguta},\ and\ \citenamefont {Blinc}}]{Zalar_prl03}%
  \BibitemOpen
  \bibfield  {author} {\bibinfo {author} {\bibfnamefont {B.~c.~v.}\
  \bibnamefont {Zalar}}, \bibinfo {author} {\bibfnamefont {V.~V.}\ \bibnamefont
  {Laguta}},\ and\ \bibinfo {author} {\bibfnamefont {R.}~\bibnamefont
  {Blinc}},\ }\bibfield  {title} {\bibinfo {title} {Nmr evidence for the
  coexistence of order-disorder and displacive components in barium titanate},\
  }\href {https://doi.org/10.1103/PhysRevLett.90.037601} {\bibfield  {journal}
  {\bibinfo  {journal} {Phys. Rev. Lett.}\ }\textbf {\bibinfo {volume} {90}},\
  \bibinfo {pages} {037601} (\bibinfo {year} {2003})}\BibitemShut {NoStop}%
\bibitem [{\citenamefont {Qi}\ \emph {et~al.}(2016)\citenamefont {Qi},
  \citenamefont {Liu}, \citenamefont {Grinberg},\ and\ \citenamefont
  {Rappe}}]{Qi_prb16}%
  \BibitemOpen
  \bibfield  {author} {\bibinfo {author} {\bibfnamefont {Y.}~\bibnamefont
  {Qi}}, \bibinfo {author} {\bibfnamefont {S.}~\bibnamefont {Liu}}, \bibinfo
  {author} {\bibfnamefont {I.}~\bibnamefont {Grinberg}},\ and\ \bibinfo
  {author} {\bibfnamefont {A.~M.}\ \bibnamefont {Rappe}},\ }\bibfield  {title}
  {\bibinfo {title} {Atomistic description for temperature-driven phase
  transitions in $\mathrm{BaTiO}_{3}$},\ }\href
  {https://doi.org/10.1103/PhysRevB.94.134308} {\bibfield  {journal} {\bibinfo
  {journal} {Phys. Rev. B}\ }\textbf {\bibinfo {volume} {94}},\ \bibinfo
  {pages} {134308} (\bibinfo {year} {2016})}\BibitemShut {NoStop}%
\bibitem [{\citenamefont {Pa\ifmmode~\acute{s}\else \'{s}\fi{}ciak}\ \emph
  {et~al.}(2018)\citenamefont {Pa\ifmmode~\acute{s}\else \'{s}\fi{}ciak},
  \citenamefont {Welberry}, \citenamefont {Kulda}, \citenamefont {Leoni},\ and\
  \citenamefont {Hlinka}}]{Pasciak_prl18}%
  \BibitemOpen
  \bibfield  {author} {\bibinfo {author} {\bibfnamefont {M.}~\bibnamefont
  {Pa\ifmmode~\acute{s}\else \'{s}\fi{}ciak}}, \bibinfo {author} {\bibfnamefont
  {T.~R.}\ \bibnamefont {Welberry}}, \bibinfo {author} {\bibfnamefont
  {J.}~\bibnamefont {Kulda}}, \bibinfo {author} {\bibfnamefont
  {S.}~\bibnamefont {Leoni}},\ and\ \bibinfo {author} {\bibfnamefont
  {J.}~\bibnamefont {Hlinka}},\ }\bibfield  {title} {\bibinfo {title} {Dynamic
  displacement disorder of cubic $\mathrm{BaTiO}_{3}$},\ }\href
  {https://doi.org/10.1103/PhysRevLett.120.167601} {\bibfield  {journal}
  {\bibinfo  {journal} {Phys. Rev. Lett.}\ }\textbf {\bibinfo {volume} {120}},\
  \bibinfo {pages} {167601} (\bibinfo {year} {2018})}\BibitemShut {NoStop}%
\bibitem [{\citenamefont {Luo}\ \emph {et~al.}(2023)\citenamefont {Luo},
  \citenamefont {Cao}, \citenamefont {Ke}, \citenamefont {Liu},\ and\
  \citenamefont {Zhou}}]{Luo_prb23}%
  \BibitemOpen
  \bibfield  {author} {\bibinfo {author} {\bibfnamefont {H.}~\bibnamefont
  {Luo}}, \bibinfo {author} {\bibfnamefont {L.}~\bibnamefont {Cao}}, \bibinfo
  {author} {\bibfnamefont {H.}~\bibnamefont {Ke}}, \bibinfo {author}
  {\bibfnamefont {G.}~\bibnamefont {Liu}},\ and\ \bibinfo {author}
  {\bibfnamefont {Y.}~\bibnamefont {Zhou}},\ }\bibfield  {title} {\bibinfo
  {title} {Statistical singularity energy in ferroelectric phase transitions},\
  }\href {https://doi.org/10.1103/PhysRevB.108.L060101} {\bibfield  {journal}
  {\bibinfo  {journal} {Phys. Rev. B}\ }\textbf {\bibinfo {volume} {108}},\
  \bibinfo {pages} {L060101} (\bibinfo {year} {2023})}\BibitemShut {NoStop}%
\bibitem [{\citenamefont {Pramanick}\ \emph {et~al.}(2015)\citenamefont
  {Pramanick}, \citenamefont {Wang}, \citenamefont {Hoffmann}, \citenamefont
  {Diallo}, \citenamefont {J\o{}rgensen},\ and\ \citenamefont
  {Wang}}]{Pramanick_prb15}%
  \BibitemOpen
  \bibfield  {author} {\bibinfo {author} {\bibfnamefont {A.}~\bibnamefont
  {Pramanick}}, \bibinfo {author} {\bibfnamefont {X.~P.}\ \bibnamefont {Wang}},
  \bibinfo {author} {\bibfnamefont {C.}~\bibnamefont {Hoffmann}}, \bibinfo
  {author} {\bibfnamefont {S.~O.}\ \bibnamefont {Diallo}}, \bibinfo {author}
  {\bibfnamefont {M.~R.~V.}\ \bibnamefont {J\o{}rgensen}},\ and\ \bibinfo
  {author} {\bibfnamefont {X.-L.}\ \bibnamefont {Wang}},\ }\bibfield  {title}
  {\bibinfo {title} {Microdomain dynamics in single-crystal
  $\mathrm{BaTiO}_{3}$ during paraelectric-ferroelectric phase transition
  measured with time-of-flight neutron scattering},\ }\href
  {https://doi.org/10.1103/PhysRevB.92.174103} {\bibfield  {journal} {\bibinfo
  {journal} {Phys. Rev. B}\ }\textbf {\bibinfo {volume} {92}},\ \bibinfo
  {pages} {174103} (\bibinfo {year} {2015})}\BibitemShut {NoStop}%
\bibitem [{\citenamefont {Zhong}\ \emph {et~al.}(1994)\citenamefont {Zhong},
  \citenamefont {Vanderbilt},\ and\ \citenamefont {Rabe}}]{Zhong_prl94}%
  \BibitemOpen
  \bibfield  {author} {\bibinfo {author} {\bibfnamefont {W.}~\bibnamefont
  {Zhong}}, \bibinfo {author} {\bibfnamefont {D.}~\bibnamefont {Vanderbilt}},\
  and\ \bibinfo {author} {\bibfnamefont {K.~M.}\ \bibnamefont {Rabe}},\
  }\bibfield  {title} {\bibinfo {title} {Phase transitions in
  $\mathrm{BaTiO}_{3}$ from first principles},\ }\href
  {https://doi.org/10.1103/PhysRevLett.73.1861} {\bibfield  {journal} {\bibinfo
   {journal} {Phys. Rev. Lett.}\ }\textbf {\bibinfo {volume} {73}},\ \bibinfo
  {pages} {1861} (\bibinfo {year} {1994})}\BibitemShut {NoStop}%
\bibitem [{\citenamefont {Pertsev}\ \emph {et~al.}(1998)\citenamefont
  {Pertsev}, \citenamefont {Zembilgotov},\ and\ \citenamefont
  {Tagantsev}}]{Pertsev_prl98}%
  \BibitemOpen
  \bibfield  {author} {\bibinfo {author} {\bibfnamefont {N.~A.}\ \bibnamefont
  {Pertsev}}, \bibinfo {author} {\bibfnamefont {A.~G.}\ \bibnamefont
  {Zembilgotov}},\ and\ \bibinfo {author} {\bibfnamefont {A.~K.}\ \bibnamefont
  {Tagantsev}},\ }\bibfield  {title} {\bibinfo {title} {Effect of mechanical
  boundary conditions on phase diagrams of epitaxial ferroelectric thin
  films},\ }\href {https://doi.org/10.1103/PhysRevLett.80.1988} {\bibfield
  {journal} {\bibinfo  {journal} {Phys. Rev. Lett.}\ }\textbf {\bibinfo
  {volume} {80}},\ \bibinfo {pages} {1988} (\bibinfo {year}
  {1998})}\BibitemShut {NoStop}%
\bibitem [{\citenamefont {Choi}\ \emph {et~al.}(2004)\citenamefont {Choi},
  \citenamefont {Biegalski}, \citenamefont {Li}, \citenamefont {Sharan},
  \citenamefont {Schubert}, \citenamefont {Uecker}, \citenamefont {Reiche},
  \citenamefont {Chen}, \citenamefont {Pan}, \citenamefont {Gopalan},
  \citenamefont {Chen}, \citenamefont {Schlom},\ and\ \citenamefont
  {Eom}}]{Choi_Science04}%
  \BibitemOpen
  \bibfield  {author} {\bibinfo {author} {\bibfnamefont {K.~J.}\ \bibnamefont
  {Choi}}, \bibinfo {author} {\bibfnamefont {M.}~\bibnamefont {Biegalski}},
  \bibinfo {author} {\bibfnamefont {Y.~L.}\ \bibnamefont {Li}}, \bibinfo
  {author} {\bibfnamefont {A.}~\bibnamefont {Sharan}}, \bibinfo {author}
  {\bibfnamefont {J.}~\bibnamefont {Schubert}}, \bibinfo {author}
  {\bibfnamefont {R.}~\bibnamefont {Uecker}}, \bibinfo {author} {\bibfnamefont
  {P.}~\bibnamefont {Reiche}}, \bibinfo {author} {\bibfnamefont {Y.~B.}\
  \bibnamefont {Chen}}, \bibinfo {author} {\bibfnamefont {X.~Q.}\ \bibnamefont
  {Pan}}, \bibinfo {author} {\bibfnamefont {V.}~\bibnamefont {Gopalan}},
  \bibinfo {author} {\bibfnamefont {L.-Q.}\ \bibnamefont {Chen}}, \bibinfo
  {author} {\bibfnamefont {D.~G.}\ \bibnamefont {Schlom}},\ and\ \bibinfo
  {author} {\bibfnamefont {C.~B.}\ \bibnamefont {Eom}},\ }\bibfield  {title}
  {\bibinfo {title} {Enhancement of ferroelectricity in strained
  $\mathrm{BaTiO}_{3}$ thin films},\ }\href
  {https://doi.org/10.1126/science.1103218} {\bibfield  {journal} {\bibinfo
  {journal} {Science}\ }\textbf {\bibinfo {volume} {306}},\ \bibinfo {pages}
  {1005} (\bibinfo {year} {2004})}\BibitemShut {NoStop}%
\bibitem [{\citenamefont {Strukov}\ \emph {et~al.}(2007)\citenamefont
  {Strukov}, \citenamefont {Davitadze}, \citenamefont {Lemanov}, \citenamefont
  {Shulman}, \citenamefont {Uesu},\ and\ \citenamefont
  {Asanuma}}]{Strukov_FerroE07}%
  \BibitemOpen
  \bibfield  {author} {\bibinfo {author} {\bibfnamefont {B.}~\bibnamefont
  {Strukov}}, \bibinfo {author} {\bibfnamefont {S.}~\bibnamefont {Davitadze}},
  \bibinfo {author} {\bibfnamefont {V.}~\bibnamefont {Lemanov}}, \bibinfo
  {author} {\bibfnamefont {S.}~\bibnamefont {Shulman}}, \bibinfo {author}
  {\bibfnamefont {Y.}~\bibnamefont {Uesu}},\ and\ \bibinfo {author}
  {\bibfnamefont {S.}~\bibnamefont {Asanuma}},\ }\bibfield  {title} {\bibinfo
  {title} {Comparative study of phase transitions in polycrystalline and
  epitaxial $\mathrm{BaTiO}_{3}$ thin films by specific heat measurements},\
  }\href@noop {} {\bibfield  {journal} {\bibinfo  {journal} {Ferroelectrics}\
  }\textbf {\bibinfo {volume} {347}},\ \bibinfo {pages} {179} (\bibinfo {year}
  {2007})}\BibitemShut {NoStop}%
\bibitem [{\citenamefont {Akcay}\ \emph {et~al.}(2008)\citenamefont {Akcay},
  \citenamefont {Alpay}, \citenamefont {Rossetti},\ and\ \citenamefont
  {Scott}}]{Scott_JAP08}%
  \BibitemOpen
  \bibfield  {author} {\bibinfo {author} {\bibfnamefont {G.}~\bibnamefont
  {Akcay}}, \bibinfo {author} {\bibfnamefont {S.~P.}\ \bibnamefont {Alpay}},
  \bibinfo {author} {\bibfnamefont {J.}~\bibnamefont {Rossetti}, \bibfnamefont
  {G.~A.}},\ and\ \bibinfo {author} {\bibfnamefont {J.~F.}\ \bibnamefont
  {Scott}},\ }\bibfield  {title} {\bibinfo {title} {Influence of mechanical
  boundary conditions on the electrocaloric properties of ferroelectric thin
  films},\ }\href {https://doi.org/10.1063/1.2831222} {\bibfield  {journal}
  {\bibinfo  {journal} {Journal of Applied Physics}\ }\textbf {\bibinfo
  {volume} {103}},\ \bibinfo {pages} {024104} (\bibinfo {year}
  {2008})}\BibitemShut {NoStop}%
\bibitem [{\citenamefont {Parker}\ \emph {et~al.}(2002)\citenamefont {Parker},
  \citenamefont {Maria},\ and\ \citenamefont {Kingon}}]{Parker_apl02}%
  \BibitemOpen
  \bibfield  {author} {\bibinfo {author} {\bibfnamefont {C.~B.}\ \bibnamefont
  {Parker}}, \bibinfo {author} {\bibfnamefont {J.-P.}\ \bibnamefont {Maria}},\
  and\ \bibinfo {author} {\bibfnamefont {A.~I.}\ \bibnamefont {Kingon}},\
  }\bibfield  {title} {\bibinfo {title} {Temperature and thickness dependent
  permittivity of $\mathrm{(Ba, Sr) TiO}_{3}$ thin films},\ }\href
  {https://doi.org/10.1063/1.1490148} {\bibfield  {journal} {\bibinfo
  {journal} {Applied Physics Letters}\ }\textbf {\bibinfo {volume} {81}},\
  \bibinfo {pages} {340} (\bibinfo {year} {2002})}\BibitemShut {NoStop}%
\bibitem [{\citenamefont {Kretschmer}\ and\ \citenamefont
  {Binder}(1979)}]{Binder_prb79}%
  \BibitemOpen
  \bibfield  {author} {\bibinfo {author} {\bibfnamefont {R.}~\bibnamefont
  {Kretschmer}}\ and\ \bibinfo {author} {\bibfnamefont {K.}~\bibnamefont
  {Binder}},\ }\bibfield  {title} {\bibinfo {title} {Surface effects on phase
  transitions in ferroelectrics and dipolar magnets},\ }\href
  {https://doi.org/10.1103/PhysRevB.20.1065} {\bibfield  {journal} {\bibinfo
  {journal} {Phys. Rev. B}\ }\textbf {\bibinfo {volume} {20}},\ \bibinfo
  {pages} {1065} (\bibinfo {year} {1979})}\BibitemShut {NoStop}%
\bibitem [{\citenamefont {Sun}\ \emph {et~al.}(2023)\citenamefont {Sun},
  \citenamefont {Gu}, \citenamefont {Li}, \citenamefont {Paudel}, \citenamefont
  {Liu}, \citenamefont {Wang}, \citenamefont {Zang}, \citenamefont {Gu},
  \citenamefont {Yang}, \citenamefont {Sun}, \citenamefont {Gu}, \citenamefont
  {Tsymbal}, \citenamefont {Liu}, \citenamefont {Huang}, \citenamefont {Wu},\
  and\ \citenamefont {Nie}}]{Sun_prl23}%
  \BibitemOpen
  \bibfield  {author} {\bibinfo {author} {\bibfnamefont {H.}~\bibnamefont
  {Sun}}, \bibinfo {author} {\bibfnamefont {J.}~\bibnamefont {Gu}}, \bibinfo
  {author} {\bibfnamefont {Y.}~\bibnamefont {Li}}, \bibinfo {author}
  {\bibfnamefont {T.~R.}\ \bibnamefont {Paudel}}, \bibinfo {author}
  {\bibfnamefont {D.}~\bibnamefont {Liu}}, \bibinfo {author} {\bibfnamefont
  {J.}~\bibnamefont {Wang}}, \bibinfo {author} {\bibfnamefont {Y.}~\bibnamefont
  {Zang}}, \bibinfo {author} {\bibfnamefont {C.}~\bibnamefont {Gu}}, \bibinfo
  {author} {\bibfnamefont {J.}~\bibnamefont {Yang}}, \bibinfo {author}
  {\bibfnamefont {W.}~\bibnamefont {Sun}}, \bibinfo {author} {\bibfnamefont
  {Z.}~\bibnamefont {Gu}}, \bibinfo {author} {\bibfnamefont {E.~Y.}\
  \bibnamefont {Tsymbal}}, \bibinfo {author} {\bibfnamefont {J.}~\bibnamefont
  {Liu}}, \bibinfo {author} {\bibfnamefont {H.}~\bibnamefont {Huang}}, \bibinfo
  {author} {\bibfnamefont {D.}~\bibnamefont {Wu}},\ and\ \bibinfo {author}
  {\bibfnamefont {Y.}~\bibnamefont {Nie}},\ }\bibfield  {title} {\bibinfo
  {title} {Prominent size effects without a depolarization field observed in
  ultrathin ferroelectric oxide membranes},\ }\href
  {https://doi.org/10.1103/PhysRevLett.130.126801} {\bibfield  {journal}
  {\bibinfo  {journal} {Phys. Rev. Lett.}\ }\textbf {\bibinfo {volume} {130}},\
  \bibinfo {pages} {126801} (\bibinfo {year} {2023})}\BibitemShut {NoStop}%
\bibitem [{\citenamefont {Catalan}\ \emph {et~al.}(2004)\citenamefont
  {Catalan}, \citenamefont {Sinnamon},\ and\ \citenamefont
  {Gregg}}]{Catalan_Jpcm04}%
  \BibitemOpen
  \bibfield  {author} {\bibinfo {author} {\bibfnamefont {G.}~\bibnamefont
  {Catalan}}, \bibinfo {author} {\bibfnamefont {L.~J.}\ \bibnamefont
  {Sinnamon}},\ and\ \bibinfo {author} {\bibfnamefont {J.~M.}\ \bibnamefont
  {Gregg}},\ }\bibfield  {title} {\bibinfo {title} {The effect of
  flexoelectricity on the dielectric properties of inhomogeneously strained
  ferroelectric thin films},\ }\href
  {https://doi.org/10.1088/0953-8984/16/13/006} {\bibfield  {journal} {\bibinfo
   {journal} {Journal of Physics: Condensed Matter}\ }\textbf {\bibinfo
  {volume} {16}},\ \bibinfo {pages} {2253} (\bibinfo {year}
  {2004})}\BibitemShut {NoStop}%
\bibitem [{\citenamefont {Yudin}\ \emph {et~al.}(2021)\citenamefont {Yudin},
  \citenamefont {Duchon}, \citenamefont {Pacherova}, \citenamefont
  {Klementova}, \citenamefont {Kocourek}, \citenamefont {Dejneka},\ and\
  \citenamefont {Tyunina}}]{Yudin_prr21}%
  \BibitemOpen
  \bibfield  {author} {\bibinfo {author} {\bibfnamefont {P.}~\bibnamefont
  {Yudin}}, \bibinfo {author} {\bibfnamefont {J.}~\bibnamefont {Duchon}},
  \bibinfo {author} {\bibfnamefont {O.}~\bibnamefont {Pacherova}}, \bibinfo
  {author} {\bibfnamefont {M.}~\bibnamefont {Klementova}}, \bibinfo {author}
  {\bibfnamefont {T.}~\bibnamefont {Kocourek}}, \bibinfo {author}
  {\bibfnamefont {A.}~\bibnamefont {Dejneka}},\ and\ \bibinfo {author}
  {\bibfnamefont {M.}~\bibnamefont {Tyunina}},\ }\bibfield  {title} {\bibinfo
  {title} {Ferroelectric phase transitions induced by a strain gradient},\
  }\href {https://doi.org/10.1103/PhysRevResearch.3.033213} {\bibfield
  {journal} {\bibinfo  {journal} {Phys. Rev. Res.}\ }\textbf {\bibinfo {volume}
  {3}},\ \bibinfo {pages} {033213} (\bibinfo {year} {2021})}\BibitemShut
  {NoStop}%
\bibitem [{\citenamefont {Imry}\ and\ \citenamefont
  {Wortis}(1979)}]{Imry_prb79}%
  \BibitemOpen
  \bibfield  {author} {\bibinfo {author} {\bibfnamefont {Y.}~\bibnamefont
  {Imry}}\ and\ \bibinfo {author} {\bibfnamefont {M.}~\bibnamefont {Wortis}},\
  }\bibfield  {title} {\bibinfo {title} {Influence of quenched impurities on
  first-order phase transitions},\ }\href
  {https://doi.org/10.1103/PhysRevB.19.3580} {\bibfield  {journal} {\bibinfo
  {journal} {Phys. Rev. B}\ }\textbf {\bibinfo {volume} {19}},\ \bibinfo
  {pages} {3580} (\bibinfo {year} {1979})}\BibitemShut {NoStop}%
\bibitem [{\citenamefont {Aizenman}\ and\ \citenamefont
  {Wehr}(1989)}]{Aizenman_prl89}%
  \BibitemOpen
  \bibfield  {author} {\bibinfo {author} {\bibfnamefont {M.}~\bibnamefont
  {Aizenman}}\ and\ \bibinfo {author} {\bibfnamefont {J.}~\bibnamefont
  {Wehr}},\ }\bibfield  {title} {\bibinfo {title} {Rounding of first-order
  phase transitions in systems with quenched disorder},\ }\href
  {https://doi.org/10.1103/PhysRevLett.62.2503} {\bibfield  {journal} {\bibinfo
   {journal} {Phys. Rev. Lett.}\ }\textbf {\bibinfo {volume} {62}},\ \bibinfo
  {pages} {2503} (\bibinfo {year} {1989})}\BibitemShut {NoStop}%
\bibitem [{\citenamefont {Cardy}\ and\ \citenamefont
  {Jacobsen}(1997)}]{Cardy_prl97}%
  \BibitemOpen
  \bibfield  {author} {\bibinfo {author} {\bibfnamefont {J.}~\bibnamefont
  {Cardy}}\ and\ \bibinfo {author} {\bibfnamefont {J.~L.}\ \bibnamefont
  {Jacobsen}},\ }\bibfield  {title} {\bibinfo {title} {Critical behavior of
  random-bond potts models},\ }\href
  {https://doi.org/10.1103/PhysRevLett.79.4063} {\bibfield  {journal} {\bibinfo
   {journal} {Phys. Rev. Lett.}\ }\textbf {\bibinfo {volume} {79}},\ \bibinfo
  {pages} {4063} (\bibinfo {year} {1997})}\BibitemShut {NoStop}%
\bibitem [{\citenamefont {Saad}\ \emph {et~al.}(2004)\citenamefont {Saad},
  \citenamefont {Baxter}, \citenamefont {Bowman}, \citenamefont {Gregg},
  \citenamefont {Morrison},\ and\ \citenamefont {Scott}}]{Saad_jpcm04}%
  \BibitemOpen
  \bibfield  {author} {\bibinfo {author} {\bibfnamefont {M.~M.}\ \bibnamefont
  {Saad}}, \bibinfo {author} {\bibfnamefont {P.}~\bibnamefont {Baxter}},
  \bibinfo {author} {\bibfnamefont {R.~M.}\ \bibnamefont {Bowman}}, \bibinfo
  {author} {\bibfnamefont {J.~M.}\ \bibnamefont {Gregg}}, \bibinfo {author}
  {\bibfnamefont {F.~D.}\ \bibnamefont {Morrison}},\ and\ \bibinfo {author}
  {\bibfnamefont {J.~F.}\ \bibnamefont {Scott}},\ }\bibfield  {title} {\bibinfo
  {title} {Intrinsic dielectric response in ferroelectric nano-capacitors},\
  }\href {https://doi.org/10.1088/0953-8984/16/41/L04} {\bibfield  {journal}
  {\bibinfo  {journal} {Journal of Physics: Condensed Matter}\ }\textbf
  {\bibinfo {volume} {16}},\ \bibinfo {pages} {L451} (\bibinfo {year}
  {2004})}\BibitemShut {NoStop}%
\bibitem [{\citenamefont {Lu}\ \emph {et~al.}(2016)\citenamefont {Lu},
  \citenamefont {Baek}, \citenamefont {Hong}, \citenamefont {Kourkoutis},
  \citenamefont {Hikita},\ and\ \citenamefont {Hwang}}]{Lu_NatMat16}%
  \BibitemOpen
  \bibfield  {author} {\bibinfo {author} {\bibfnamefont {D.}~\bibnamefont
  {Lu}}, \bibinfo {author} {\bibfnamefont {D.~J.}\ \bibnamefont {Baek}},
  \bibinfo {author} {\bibfnamefont {S.~S.}\ \bibnamefont {Hong}}, \bibinfo
  {author} {\bibfnamefont {L.~F.}\ \bibnamefont {Kourkoutis}}, \bibinfo
  {author} {\bibfnamefont {Y.}~\bibnamefont {Hikita}},\ and\ \bibinfo {author}
  {\bibfnamefont {H.~Y.}\ \bibnamefont {Hwang}},\ }\bibfield  {title} {\bibinfo
  {title} {Synthesis of freestanding single-crystal perovskite films and
  heterostructures by etching of sacrificial water-soluble layers. nature
  materials},\ }\href {https://doi.org/10.1038/nmat4749} {\bibfield  {journal}
  {\bibinfo  {journal} {Nature materials}\ }\textbf {\bibinfo {volume} {15}},\
  \bibinfo {pages} {1255} (\bibinfo {year} {2016})}\BibitemShut {NoStop}%
\bibitem [{\citenamefont {Pesquera}\ \emph {et~al.}(2020)\citenamefont
  {Pesquera}, \citenamefont {Parsonnet}, \citenamefont {Qualls}, \citenamefont
  {Xu}, \citenamefont {Gubser}, \citenamefont {Kim}, \citenamefont {Jiang},
  \citenamefont {Velarde}, \citenamefont {Huang}, \citenamefont {Hwang},
  \citenamefont {Ramesh},\ and\ \citenamefont {Martin}}]{David_AdvMat20}%
  \BibitemOpen
  \bibfield  {author} {\bibinfo {author} {\bibfnamefont {D.}~\bibnamefont
  {Pesquera}}, \bibinfo {author} {\bibfnamefont {E.}~\bibnamefont {Parsonnet}},
  \bibinfo {author} {\bibfnamefont {A.}~\bibnamefont {Qualls}}, \bibinfo
  {author} {\bibfnamefont {R.}~\bibnamefont {Xu}}, \bibinfo {author}
  {\bibfnamefont {A.~J.}\ \bibnamefont {Gubser}}, \bibinfo {author}
  {\bibfnamefont {J.}~\bibnamefont {Kim}}, \bibinfo {author} {\bibfnamefont
  {Y.}~\bibnamefont {Jiang}}, \bibinfo {author} {\bibfnamefont
  {G.}~\bibnamefont {Velarde}}, \bibinfo {author} {\bibfnamefont {Y.-L.}\
  \bibnamefont {Huang}}, \bibinfo {author} {\bibfnamefont {H.~Y.}\ \bibnamefont
  {Hwang}}, \bibinfo {author} {\bibfnamefont {R.}~\bibnamefont {Ramesh}},\ and\
  \bibinfo {author} {\bibfnamefont {L.~W.}\ \bibnamefont {Martin}},\ }\bibfield
   {title} {\bibinfo {title} {Beyond substrates: Strain engineering of
  ferroelectric membranes},\ }\href
  {https://doi.org/https://doi.org/10.1002/adma.202003780} {\bibfield
  {journal} {\bibinfo  {journal} {Advanced Materials}\ }\textbf {\bibinfo
  {volume} {32}},\ \bibinfo {pages} {2003780} (\bibinfo {year}
  {2020})}\BibitemShut {NoStop}%
\bibitem [{\citenamefont {Pesquera}\ \emph {et~al.}(2022)\citenamefont
  {Pesquera}, \citenamefont {Fernández}, \citenamefont {Khestanova},\ and\
  \citenamefont {Martin}}]{Pesquera_JPCM22}%
  \BibitemOpen
  \bibfield  {author} {\bibinfo {author} {\bibfnamefont {D.}~\bibnamefont
  {Pesquera}}, \bibinfo {author} {\bibfnamefont {A.}~\bibnamefont
  {Fernández}}, \bibinfo {author} {\bibfnamefont {E.}~\bibnamefont
  {Khestanova}},\ and\ \bibinfo {author} {\bibfnamefont {L.~W.}\ \bibnamefont
  {Martin}},\ }\bibfield  {title} {\bibinfo {title} {Freestanding complex-oxide
  membranes},\ }\href {https://doi.org/10.1088/1361-648X/ac7dd5} {\bibfield
  {journal} {\bibinfo  {journal} {Journal of Physics: Condensed Matter}\
  }\textbf {\bibinfo {volume} {34}},\ \bibinfo {pages} {383001} (\bibinfo
  {year} {2022})}\BibitemShut {NoStop}%
\bibitem [{\citenamefont {Ganguly}\ \emph {et~al.}(2024)\citenamefont
  {Ganguly}, \citenamefont {Pesquera}, \citenamefont {Garcia}, \citenamefont
  {Saeed}, \citenamefont {Mirzamohammadi}, \citenamefont {Santiso},
  \citenamefont {Padilla}, \citenamefont {Roque}, \citenamefont {Laulhé},
  \citenamefont {Berenguer}, \citenamefont {Villanueva},\ and\ \citenamefont
  {Catalan}}]{Ganguly_AdvMat24}%
  \BibitemOpen
  \bibfield  {author} {\bibinfo {author} {\bibfnamefont {S.}~\bibnamefont
  {Ganguly}}, \bibinfo {author} {\bibfnamefont {D.}~\bibnamefont {Pesquera}},
  \bibinfo {author} {\bibfnamefont {D.~M.}\ \bibnamefont {Garcia}}, \bibinfo
  {author} {\bibfnamefont {U.}~\bibnamefont {Saeed}}, \bibinfo {author}
  {\bibfnamefont {N.}~\bibnamefont {Mirzamohammadi}}, \bibinfo {author}
  {\bibfnamefont {J.}~\bibnamefont {Santiso}}, \bibinfo {author} {\bibfnamefont
  {J.}~\bibnamefont {Padilla}}, \bibinfo {author} {\bibfnamefont {J.~M.~C.}\
  \bibnamefont {Roque}}, \bibinfo {author} {\bibfnamefont {C.}~\bibnamefont
  {Laulhé}}, \bibinfo {author} {\bibfnamefont {F.}~\bibnamefont {Berenguer}},
  \bibinfo {author} {\bibfnamefont {L.~G.}\ \bibnamefont {Villanueva}},\ and\
  \bibinfo {author} {\bibfnamefont {G.}~\bibnamefont {Catalan}},\ }\bibfield
  {title} {\bibinfo {title} {Photostrictive actuators based on freestanding
  ferroelectric membranes},\ }\href {https://doi.org/10.1002/adma.202310198}
  {\bibfield  {journal} {\bibinfo  {journal} {Advanced Materials}\ }\textbf
  {\bibinfo {volume} {36}},\ \bibinfo {pages} {2310198} (\bibinfo {year}
  {2024})}\BibitemShut {NoStop}%
\bibitem [{\citenamefont {Lee}\ \emph {et~al.}(2022)\citenamefont {Lee},
  \citenamefont {Renshof}, \citenamefont {van Zeggeren}, \citenamefont
  {Houmes}, \citenamefont {Lesne}, \citenamefont {Šiškins}, \citenamefont
  {van Thiel}, \citenamefont {Guis}, \citenamefont {van Blankenstein},
  \citenamefont {Verbiest}, \citenamefont {Caviglia}, \citenamefont {van~der
  Zant},\ and\ \citenamefont {Steeneken}}]{Lee_AdvMat22}%
  \BibitemOpen
  \bibfield  {author} {\bibinfo {author} {\bibfnamefont {M.}~\bibnamefont
  {Lee}}, \bibinfo {author} {\bibfnamefont {J.~R.}\ \bibnamefont {Renshof}},
  \bibinfo {author} {\bibfnamefont {K.~J.}\ \bibnamefont {van Zeggeren}},
  \bibinfo {author} {\bibfnamefont {M.~J.~A.}\ \bibnamefont {Houmes}}, \bibinfo
  {author} {\bibfnamefont {E.}~\bibnamefont {Lesne}}, \bibinfo {author}
  {\bibfnamefont {M.}~\bibnamefont {Šiškins}}, \bibinfo {author}
  {\bibfnamefont {T.~C.}\ \bibnamefont {van Thiel}}, \bibinfo {author}
  {\bibfnamefont {R.~H.}\ \bibnamefont {Guis}}, \bibinfo {author}
  {\bibfnamefont {M.~R.}\ \bibnamefont {van Blankenstein}}, \bibinfo {author}
  {\bibfnamefont {G.~J.}\ \bibnamefont {Verbiest}}, \bibinfo {author}
  {\bibfnamefont {A.~D.}\ \bibnamefont {Caviglia}}, \bibinfo {author}
  {\bibfnamefont {H.~S.~J.}\ \bibnamefont {van~der Zant}},\ and\ \bibinfo
  {author} {\bibfnamefont {P.~G.}\ \bibnamefont {Steeneken}},\ }\bibfield
  {title} {\bibinfo {title} {Ultrathin piezoelectric resonators based on
  graphene and free-standing single-crystal $\mathrm{BaTiO}_{3}$},\ }\href
  {https://doi.org/10.1002/adma.202204630} {\bibfield  {journal} {\bibinfo
  {journal} {Advanced Materials}\ }\textbf {\bibinfo {volume} {34}},\ \bibinfo
  {pages} {2204630} (\bibinfo {year} {2022})}\BibitemShut {NoStop}%
\bibitem [{\citenamefont {Efremov}\ \emph {et~al.}(2004)\citenamefont
  {Efremov}, \citenamefont {Olson}, \citenamefont {Zhang}, \citenamefont {Lai},
  \citenamefont {Schiettekatte}, \citenamefont {Zhang},\ and\ \citenamefont
  {Allen}}]{Allen_ThermoActa04}%
  \BibitemOpen
  \bibfield  {author} {\bibinfo {author} {\bibfnamefont {M.}~\bibnamefont
  {Efremov}}, \bibinfo {author} {\bibfnamefont {E.}~\bibnamefont {Olson}},
  \bibinfo {author} {\bibfnamefont {M.}~\bibnamefont {Zhang}}, \bibinfo
  {author} {\bibfnamefont {S.}~\bibnamefont {Lai}}, \bibinfo {author}
  {\bibfnamefont {F.}~\bibnamefont {Schiettekatte}}, \bibinfo {author}
  {\bibfnamefont {Z.}~\bibnamefont {Zhang}},\ and\ \bibinfo {author}
  {\bibfnamefont {L.}~\bibnamefont {Allen}},\ }\bibfield  {title} {\bibinfo
  {title} {Thin-film differential scanning nanocalorimetry: heat capacity
  analysis},\ }\href
  {https://doi.org/https://doi.org/10.1016/j.tca.2003.08.019} {\bibfield
  {journal} {\bibinfo  {journal} {Thermochimica Acta}\ }\textbf {\bibinfo
  {volume} {412}},\ \bibinfo {pages} {13} (\bibinfo {year} {2004})}\BibitemShut
  {NoStop}%
\bibitem [{\citenamefont {Rodr{\'i}guez-Viejo}\ and\ \citenamefont
  {Lopeand{\'i}a}(2016)}]{Rodríguez-Viejo_Book16}%
  \BibitemOpen
  \bibfield  {author} {\bibinfo {author} {\bibfnamefont {J.}~\bibnamefont
  {Rodr{\'i}guez-Viejo}}\ and\ \bibinfo {author} {\bibfnamefont {A.~F.}\
  \bibnamefont {Lopeand{\'i}a}},\ }\bibinfo {title} {Quasi-adiabatic,
  membrane-based, highly sensitive fast scanning nanocalorimetry},\ in\ \href
  {https://doi.org/10.1007/978-3-319-31329-0_3} {\emph {\bibinfo {booktitle}
  {Fast Scanning Calorimetry}}},\ \bibinfo {editor} {edited by\ \bibinfo
  {editor} {\bibfnamefont {C.}~\bibnamefont {Schick}}\ and\ \bibinfo {editor}
  {\bibfnamefont {V.}~\bibnamefont {Mathot}}}\ (\bibinfo  {publisher} {Springer
  International Publishing},\ \bibinfo {address} {Cham},\ \bibinfo {year}
  {2016})\ pp.\ \bibinfo {pages} {105--149}\BibitemShut {NoStop}%
\bibitem [{\citenamefont {Vila-Costa}\ \emph {et~al.}(2020)\citenamefont
  {Vila-Costa}, \citenamefont {R\`afols-Rib\'e}, \citenamefont
  {Gonz\'alez-Silveira}, \citenamefont {Lopeandia}, \citenamefont
  {Abad-Mu\~noz},\ and\ \citenamefont
  {Rodr\'{\i}guez-Viejo}}]{Vila-Costa_prl20}%
  \BibitemOpen
  \bibfield  {author} {\bibinfo {author} {\bibfnamefont {A.}~\bibnamefont
  {Vila-Costa}}, \bibinfo {author} {\bibfnamefont {J.}~\bibnamefont
  {R\`afols-Rib\'e}}, \bibinfo {author} {\bibfnamefont {M.}~\bibnamefont
  {Gonz\'alez-Silveira}}, \bibinfo {author} {\bibfnamefont {A.~F.}\
  \bibnamefont {Lopeandia}}, \bibinfo {author} {\bibfnamefont {L.}~\bibnamefont
  {Abad-Mu\~noz}},\ and\ \bibinfo {author} {\bibfnamefont {J.}~\bibnamefont
  {Rodr\'{\i}guez-Viejo}},\ }\bibfield  {title} {\bibinfo {title} {Nucleation
  and growth of the supercooled liquid phase control glass transition in bulk
  ultrastable glasses},\ }\href
  {https://doi.org/10.1103/PhysRevLett.124.076002} {\bibfield  {journal}
  {\bibinfo  {journal} {Phys. Rev. Lett.}\ }\textbf {\bibinfo {volume} {124}},\
  \bibinfo {pages} {076002} (\bibinfo {year} {2020})}\BibitemShut {NoStop}%
\bibitem [{sup()}]{supplementary}%
  \BibitemOpen
  \href@noop {} {\emph {\bibinfo {title} {See the Supplemental Material for
  details of the experiments and data analysis. The Supplemental Material
  includes references \cite{Allen_ThermoActa04, Rodríguez-Viejo_Book16,
  Vila-Costa_prl20, Javier_prb11, Lu_NatMat16, David_AdvMat20, Pesquera_JPCM22,
  Mathur_AdvMat13}.}}}\BibitemShut {Stop}%
\bibitem [{\citenamefont {Rodriguez}\ \emph {et~al.}(2007)\citenamefont
  {Rodriguez}, \citenamefont {Callahan}, \citenamefont {Kalinin},\ and\
  \citenamefont {Proksch}}]{Rodriguez_NanoTech2007}%
  \BibitemOpen
  \bibfield  {author} {\bibinfo {author} {\bibfnamefont {B.~J.}\ \bibnamefont
  {Rodriguez}}, \bibinfo {author} {\bibfnamefont {C.}~\bibnamefont {Callahan}},
  \bibinfo {author} {\bibfnamefont {S.~V.}\ \bibnamefont {Kalinin}},\ and\
  \bibinfo {author} {\bibfnamefont {R.}~\bibnamefont {Proksch}},\ }\bibfield
  {title} {\bibinfo {title} {Dual-frequency resonance-tracking atomic force
  microscopy},\ }\href {https://doi.org/10.1088/0957-4484/18/47/475504}
  {\bibfield  {journal} {\bibinfo  {journal} {Nanotechnology}\ }\textbf
  {\bibinfo {volume} {18}},\ \bibinfo {pages} {475504} (\bibinfo {year}
  {2007})}\BibitemShut {NoStop}%
\bibitem [{\citenamefont {Rabe}\ \emph {et~al.}(2007)\citenamefont {Rabe},
  \citenamefont {Dawber}, \citenamefont {Lichtensteiger}, \citenamefont {Ahn},\
  and\ \citenamefont {Triscone}}]{Book_Rabe07}%
  \BibitemOpen
  \bibfield  {author} {\bibinfo {author} {\bibfnamefont {K.~M.}\ \bibnamefont
  {Rabe}}, \bibinfo {author} {\bibfnamefont {M.}~\bibnamefont {Dawber}},
  \bibinfo {author} {\bibfnamefont {C.}~\bibnamefont {Lichtensteiger}},
  \bibinfo {author} {\bibfnamefont {C.~H.}\ \bibnamefont {Ahn}},\ and\ \bibinfo
  {author} {\bibfnamefont {J.-M.}\ \bibnamefont {Triscone}},\ }\bibfield
  {title} {\bibinfo {title} {Modern physics of ferroelectrics: Essential
  background},\ }in\ \href@noop {} {\emph {\bibinfo {booktitle} {Physics of
  Ferroelectrics: A Modern Perspective}}},\ Vol.\ \bibinfo {volume} {105}\
  (\bibinfo  {publisher} {Springer},\ \bibinfo {year} {2007})\BibitemShut
  {NoStop}%
\bibitem [{\citenamefont {Chrosch}\ and\ \citenamefont
  {Salje}(1999)}]{Chrosch_jap99}%
  \BibitemOpen
  \bibfield  {author} {\bibinfo {author} {\bibfnamefont {J.}~\bibnamefont
  {Chrosch}}\ and\ \bibinfo {author} {\bibfnamefont {E.~K.~H.}\ \bibnamefont
  {Salje}},\ }\bibfield  {title} {\bibinfo {title} {Temperature dependence of
  the domain wall width in $\mathrm{LaAlO}_{3}$},\ }\href
  {https://doi.org/10.1063/1.369152} {\bibfield  {journal} {\bibinfo  {journal}
  {Journal of Applied Physics}\ }\textbf {\bibinfo {volume} {85}},\ \bibinfo
  {pages} {722} (\bibinfo {year} {1999})}\BibitemShut {NoStop}%
\bibitem [{\citenamefont {Shih}\ \emph {et~al.}(1994)\citenamefont {Shih},
  \citenamefont {Shih},\ and\ \citenamefont {Aksay}}]{Shih_prb94}%
  \BibitemOpen
  \bibfield  {author} {\bibinfo {author} {\bibfnamefont {W.~Y.}\ \bibnamefont
  {Shih}}, \bibinfo {author} {\bibfnamefont {W.-H.}\ \bibnamefont {Shih}},\
  and\ \bibinfo {author} {\bibfnamefont {I.~A.}\ \bibnamefont {Aksay}},\
  }\bibfield  {title} {\bibinfo {title} {Size dependence of the ferroelectric
  transition of small ${\mathrm{batio}}_{3}$ particles: Effect of
  depolarization},\ }\href {https://doi.org/10.1103/PhysRevB.50.15575}
  {\bibfield  {journal} {\bibinfo  {journal} {Phys. Rev. B}\ }\textbf {\bibinfo
  {volume} {50}},\ \bibinfo {pages} {15575} (\bibinfo {year}
  {1994})}\BibitemShut {NoStop}%
\bibitem [{\citenamefont {Stephanovich}\ \emph {et~al.}(2003)\citenamefont
  {Stephanovich}, \citenamefont {Luk'Yanchuk},\ and\ \citenamefont
  {Karkut}}]{Stephanovich_Ferroelectics03}%
  \BibitemOpen
  \bibfield  {author} {\bibinfo {author} {\bibfnamefont {V.~A.}\ \bibnamefont
  {Stephanovich}}, \bibinfo {author} {\bibfnamefont {I.~A.}\ \bibnamefont
  {Luk'Yanchuk}},\ and\ \bibinfo {author} {\bibfnamefont {M.~G.}\ \bibnamefont
  {Karkut}},\ }\bibfield  {title} {\bibinfo {title} {Domain proximity and
  ferroelectric transition in ferro-paraelectric superlattices},\ }\href
  {https://doi.org/10.1080/00150190390222664} {\bibfield  {journal} {\bibinfo
  {journal} {Ferroelectrics}\ }\textbf {\bibinfo {volume} {291}},\ \bibinfo
  {pages} {169} (\bibinfo {year} {2003})}\BibitemShut {NoStop}%
\bibitem [{\citenamefont {Stephanovich}\ \emph {et~al.}(2005)\citenamefont
  {Stephanovich}, \citenamefont {Luk'yanchuk},\ and\ \citenamefont
  {Karkut}}]{Stephanovich_prl05}%
  \BibitemOpen
  \bibfield  {author} {\bibinfo {author} {\bibfnamefont {V.~A.}\ \bibnamefont
  {Stephanovich}}, \bibinfo {author} {\bibfnamefont {I.~A.}\ \bibnamefont
  {Luk'yanchuk}},\ and\ \bibinfo {author} {\bibfnamefont {M.~G.}\ \bibnamefont
  {Karkut}},\ }\bibfield  {title} {\bibinfo {title} {Domain-enhanced interlayer
  coupling in ferroelectric/paraelectric superlattices},\ }\href
  {https://doi.org/10.1103/PhysRevLett.94.047601} {\bibfield  {journal}
  {\bibinfo  {journal} {Phys. Rev. Lett.}\ }\textbf {\bibinfo {volume} {94}},\
  \bibinfo {pages} {047601} (\bibinfo {year} {2005})}\BibitemShut {NoStop}%
\bibitem [{\citenamefont {Luk'yanchuk}\ \emph
  {et~al.}(2009{\natexlab{a}})\citenamefont {Luk'yanchuk}, \citenamefont
  {Lahoche},\ and\ \citenamefont {Sen\'e}}]{Lukyanchuk_prl09}%
  \BibitemOpen
  \bibfield  {author} {\bibinfo {author} {\bibfnamefont {I.~A.}\ \bibnamefont
  {Luk'yanchuk}}, \bibinfo {author} {\bibfnamefont {L.}~\bibnamefont
  {Lahoche}},\ and\ \bibinfo {author} {\bibfnamefont {A.}~\bibnamefont
  {Sen\'e}},\ }\bibfield  {title} {\bibinfo {title} {Universal properties of
  ferroelectric domains},\ }\href
  {https://doi.org/10.1103/PhysRevLett.102.147601} {\bibfield  {journal}
  {\bibinfo  {journal} {Phys. Rev. Lett.}\ }\textbf {\bibinfo {volume} {102}},\
  \bibinfo {pages} {147601} (\bibinfo {year} {2009}{\natexlab{a}})}\BibitemShut
  {NoStop}%
\bibitem [{\citenamefont {Bratkovsky}\ and\ \citenamefont
  {Levanyuk}(2000)}]{Bratkovsky_prl00}%
  \BibitemOpen
  \bibfield  {author} {\bibinfo {author} {\bibfnamefont {A.~M.}\ \bibnamefont
  {Bratkovsky}}\ and\ \bibinfo {author} {\bibfnamefont {A.~P.}\ \bibnamefont
  {Levanyuk}},\ }\bibfield  {title} {\bibinfo {title} {Abrupt appearance of the
  domain pattern and fatigue of thin ferroelectric films},\ }\href
  {https://doi.org/10.1103/PhysRevLett.84.3177} {\bibfield  {journal} {\bibinfo
   {journal} {Phys. Rev. Lett.}\ }\textbf {\bibinfo {volume} {84}},\ \bibinfo
  {pages} {3177} (\bibinfo {year} {2000})}\BibitemShut {NoStop}%
\bibitem [{\citenamefont {Zubko}\ \emph {et~al.}(2010)\citenamefont {Zubko},
  \citenamefont {Stucki}, \citenamefont {Lichtensteiger},\ and\ \citenamefont
  {Triscone}}]{Zubko_prl10}%
  \BibitemOpen
  \bibfield  {author} {\bibinfo {author} {\bibfnamefont {P.}~\bibnamefont
  {Zubko}}, \bibinfo {author} {\bibfnamefont {N.}~\bibnamefont {Stucki}},
  \bibinfo {author} {\bibfnamefont {C.}~\bibnamefont {Lichtensteiger}},\ and\
  \bibinfo {author} {\bibfnamefont {J.-M.}\ \bibnamefont {Triscone}},\
  }\bibfield  {title} {\bibinfo {title} {X-ray diffraction studies of
  180\ifmmode^\circ\else\textdegree\fi{} ferroelectric domains in
  ${\mathrm{pbtio}}_{3}/{\mathrm{srtio}}_{3}$ superlattices under an applied
  electric field},\ }\href {https://doi.org/10.1103/PhysRevLett.104.187601}
  {\bibfield  {journal} {\bibinfo  {journal} {Phys. Rev. Lett.}\ }\textbf
  {\bibinfo {volume} {104}},\ \bibinfo {pages} {187601} (\bibinfo {year}
  {2010})}\BibitemShut {NoStop}%
\bibitem [{\citenamefont {Kornev}\ \emph {et~al.}(2004)\citenamefont {Kornev},
  \citenamefont {Fu},\ and\ \citenamefont {Bellaiche}}]{Kornev_prl04}%
  \BibitemOpen
  \bibfield  {author} {\bibinfo {author} {\bibfnamefont {I.}~\bibnamefont
  {Kornev}}, \bibinfo {author} {\bibfnamefont {H.}~\bibnamefont {Fu}},\ and\
  \bibinfo {author} {\bibfnamefont {L.}~\bibnamefont {Bellaiche}},\ }\bibfield
  {title} {\bibinfo {title} {Ultrathin films of ferroelectric solid solutions
  under a residual depolarizing field},\ }\href
  {https://doi.org/10.1103/PhysRevLett.93.196104} {\bibfield  {journal}
  {\bibinfo  {journal} {Phys. Rev. Lett.}\ }\textbf {\bibinfo {volume} {93}},\
  \bibinfo {pages} {196104} (\bibinfo {year} {2004})}\BibitemShut {NoStop}%
\bibitem [{\citenamefont {Yadav}\ \emph {et~al.}(2016)\citenamefont {Yadav},
  \citenamefont {Nelson}, \citenamefont {Hsu}, \citenamefont {Hong},
  \citenamefont {Clarkson}, \citenamefont {Schlep{\"u}tz}, \citenamefont
  {Damodaran}, \citenamefont {Shafer}, \citenamefont {Arenholz}, \citenamefont
  {Dedon}, \citenamefont {Chen}, \citenamefont {Vishwanath}, \citenamefont
  {Minor}, \citenamefont {Chen}, \citenamefont {Scott}, \citenamefont
  {Martin},\ and\ \citenamefont {Ramesh}}]{Yadav_Nature16}%
  \BibitemOpen
  \bibfield  {author} {\bibinfo {author} {\bibfnamefont {A.~K.}\ \bibnamefont
  {Yadav}}, \bibinfo {author} {\bibfnamefont {C.~T.}\ \bibnamefont {Nelson}},
  \bibinfo {author} {\bibfnamefont {S.~L.}\ \bibnamefont {Hsu}}, \bibinfo
  {author} {\bibfnamefont {Z.}~\bibnamefont {Hong}}, \bibinfo {author}
  {\bibfnamefont {J.~D.}\ \bibnamefont {Clarkson}}, \bibinfo {author}
  {\bibfnamefont {C.~M.}\ \bibnamefont {Schlep{\"u}tz}}, \bibinfo {author}
  {\bibfnamefont {A.~R.}\ \bibnamefont {Damodaran}}, \bibinfo {author}
  {\bibfnamefont {P.}~\bibnamefont {Shafer}}, \bibinfo {author} {\bibfnamefont
  {E.}~\bibnamefont {Arenholz}}, \bibinfo {author} {\bibfnamefont {L.~R.}\
  \bibnamefont {Dedon}}, \bibinfo {author} {\bibfnamefont {D.}~\bibnamefont
  {Chen}}, \bibinfo {author} {\bibfnamefont {A.}~\bibnamefont {Vishwanath}},
  \bibinfo {author} {\bibfnamefont {A.~M.}\ \bibnamefont {Minor}}, \bibinfo
  {author} {\bibfnamefont {L.~Q.}\ \bibnamefont {Chen}}, \bibinfo {author}
  {\bibfnamefont {J.~F.}\ \bibnamefont {Scott}}, \bibinfo {author}
  {\bibfnamefont {L.~W.}\ \bibnamefont {Martin}},\ and\ \bibinfo {author}
  {\bibfnamefont {R.}~\bibnamefont {Ramesh}},\ }\bibfield  {title} {\bibinfo
  {title} {Observation of polar vortices in oxide superlattices},\ }\href
  {https://doi.org/10.1038/nature16463} {\bibfield  {journal} {\bibinfo
  {journal} {Nature}\ }\textbf {\bibinfo {volume} {530}},\ \bibinfo {pages}
  {198} (\bibinfo {year} {2016})}\BibitemShut {NoStop}%
\bibitem [{\citenamefont {Luk'yanchuk}\ \emph {et~al.}(2024)\citenamefont
  {Luk'yanchuk}, \citenamefont {Razumnaya}, \citenamefont {Kondovych},
  \citenamefont {Tikhonov},\ and\ \citenamefont
  {Vinokur}}]{Lukyanchuk_arXiv24}%
  \BibitemOpen
  \bibfield  {author} {\bibinfo {author} {\bibfnamefont {I.}~\bibnamefont
  {Luk'yanchuk}}, \bibinfo {author} {\bibfnamefont {A.}~\bibnamefont
  {Razumnaya}}, \bibinfo {author} {\bibfnamefont {S.}~\bibnamefont
  {Kondovych}}, \bibinfo {author} {\bibfnamefont {Y.}~\bibnamefont
  {Tikhonov}},\ and\ \bibinfo {author} {\bibfnamefont {V.~M.}\ \bibnamefont
  {Vinokur}},\ }\bibfield  {title} {\bibinfo {title} {Topological ferroelectric
  chirality},\ }\href@noop {} {\bibfield  {journal} {\bibinfo  {journal} {arXiv
  preprint arXiv:2406.19728}\ } (\bibinfo {year} {2024})}\BibitemShut {NoStop}%
\bibitem [{Luk()}]{LukyanchukBoron2026Zenodo}%
  \BibitemOpen
  \href@noop {} {\emph {\bibinfo {title} {MATLAB code for the evaluation of
  fourth-order coefficient B:
  https://doi.org/10.5281/zenodo.19711781}}}\BibitemShut {NoStop}%
\bibitem [{\citenamefont {Kondovych}\ \emph {et~al.}(2025)\citenamefont
  {Kondovych}, \citenamefont {Boron}, \citenamefont {Di~Rino}, \citenamefont
  {Sepliarsky}, \citenamefont {Razumnaya}, \citenamefont {Sen{\'e}},\ and\
  \citenamefont {Lukyanchuk}}]{Kondovych2025}%
  \BibitemOpen
  \bibfield  {author} {\bibinfo {author} {\bibfnamefont {S.}~\bibnamefont
  {Kondovych}}, \bibinfo {author} {\bibfnamefont {L.}~\bibnamefont {Boron}},
  \bibinfo {author} {\bibfnamefont {F.~N.}\ \bibnamefont {Di~Rino}}, \bibinfo
  {author} {\bibfnamefont {M.}~\bibnamefont {Sepliarsky}}, \bibinfo {author}
  {\bibfnamefont {A.~G.}\ \bibnamefont {Razumnaya}}, \bibinfo {author}
  {\bibfnamefont {A.}~\bibnamefont {Sen{\'e}}},\ and\ \bibinfo {author}
  {\bibfnamefont {I.~A.}\ \bibnamefont {Lukyanchuk}},\ }\bibfield  {title}
  {\bibinfo {title} {Surface-tension-induced phase transitions in freestanding
  ferroelectric thin films},\ }\href
  {https://doi.org/10.1021/acs.nanolett.5c03216} {\bibfield  {journal}
  {\bibinfo  {journal} {Nano Letters}\ }\textbf {\bibinfo {volume} {25}},\
  \bibinfo {pages} {12987} (\bibinfo {year} {2025})},\ \bibinfo {note} {pMID:
  40808329}\BibitemShut {NoStop}%
\bibitem [{\citenamefont {Uchino}\ \emph {et~al.}(1989)\citenamefont {Uchino},
  \citenamefont {Sadanaga},\ and\ \citenamefont {Hirose}}]{Uchino1989}%
  \BibitemOpen
  \bibfield  {author} {\bibinfo {author} {\bibfnamefont {K.}~\bibnamefont
  {Uchino}}, \bibinfo {author} {\bibfnamefont {E.}~\bibnamefont {Sadanaga}},\
  and\ \bibinfo {author} {\bibfnamefont {T.}~\bibnamefont {Hirose}},\
  }\bibfield  {title} {\bibinfo {title} {Dependence of the crystal structure on
  particle size in barium titanate},\ }\href
  {https://doi.org/https://doi.org/10.1111/j.1151-2916.1989.tb07706.x}
  {\bibfield  {journal} {\bibinfo  {journal} {Journal of the American Ceramic
  Society}\ }\textbf {\bibinfo {volume} {72}},\ \bibinfo {pages} {1555}
  (\bibinfo {year} {1989})}\BibitemShut {NoStop}%
\bibitem [{\citenamefont {Cammarata}(1994)}]{Cammarata1994}%
  \BibitemOpen
  \bibfield  {author} {\bibinfo {author} {\bibfnamefont {R.~C.}\ \bibnamefont
  {Cammarata}},\ }\bibfield  {title} {\bibinfo {title} {Surface and interface
  stress effects in thin films},\ }\href
  {https://doi.org/https://doi.org/10.1016/0079-6816(94)90005-1} {\bibfield
  {journal} {\bibinfo  {journal} {Progress in Surface Science}\ }\textbf
  {\bibinfo {volume} {46}},\ \bibinfo {pages} {1} (\bibinfo {year}
  {1994})}\BibitemShut {NoStop}%
\bibitem [{\citenamefont {Ma}(2009)}]{Ma2009}%
  \BibitemOpen
  \bibfield  {author} {\bibinfo {author} {\bibfnamefont {W.}~\bibnamefont
  {Ma}},\ }\bibfield  {title} {\bibinfo {title} {Surface tension and curie
  temperature in ferroelectric nanowires and nanodots},\ }\href
  {https://doi.org/10.1007/s00339-009-5246-7} {\bibfield  {journal} {\bibinfo
  {journal} {Applied Physics A}\ }\textbf {\bibinfo {volume} {96}},\ \bibinfo
  {pages} {915} (\bibinfo {year} {2009})}\BibitemShut {NoStop}%
\bibitem [{\citenamefont {Diehm}\ \emph {et~al.}(2012)\citenamefont {Diehm},
  \citenamefont {Ágoston},\ and\ \citenamefont {Albe}}]{Diehm2012}%
  \BibitemOpen
  \bibfield  {author} {\bibinfo {author} {\bibfnamefont {P.~M.}\ \bibnamefont
  {Diehm}}, \bibinfo {author} {\bibfnamefont {P.}~\bibnamefont {Ágoston}},\
  and\ \bibinfo {author} {\bibfnamefont {K.}~\bibnamefont {Albe}},\ }\bibfield
  {title} {\bibinfo {title} {Size-dependent lattice expansion in nanoparticles:
  Reality or anomaly?},\ }\href
  {https://doi.org/https://doi.org/10.1002/cphc.201200257} {\bibfield
  {journal} {\bibinfo  {journal} {ChemPhysChem}\ }\textbf {\bibinfo {volume}
  {13}},\ \bibinfo {pages} {2443} (\bibinfo {year} {2012})}\BibitemShut
  {NoStop}%
\bibitem [{\citenamefont {Luk'yanchuk}\ \emph
  {et~al.}(2009{\natexlab{b}})\citenamefont {Luk'yanchuk}, \citenamefont
  {Schilling}, \citenamefont {Gregg}, \citenamefont {Catalan},\ and\
  \citenamefont {Scott}}]{Lukyanchuk_prb09}%
  \BibitemOpen
  \bibfield  {author} {\bibinfo {author} {\bibfnamefont {I.~A.}\ \bibnamefont
  {Luk'yanchuk}}, \bibinfo {author} {\bibfnamefont {A.}~\bibnamefont
  {Schilling}}, \bibinfo {author} {\bibfnamefont {J.~M.}\ \bibnamefont
  {Gregg}}, \bibinfo {author} {\bibfnamefont {G.}~\bibnamefont {Catalan}},\
  and\ \bibinfo {author} {\bibfnamefont {J.~F.}\ \bibnamefont {Scott}},\
  }\bibfield  {title} {\bibinfo {title} {Origin of ferroelastic domains in
  free-standing single-crystal ferroelectric films},\ }\href
  {https://doi.org/10.1103/PhysRevB.79.144111} {\bibfield  {journal} {\bibinfo
  {journal} {Phys. Rev. B}\ }\textbf {\bibinfo {volume} {79}},\ \bibinfo
  {pages} {144111} (\bibinfo {year} {2009}{\natexlab{b}})}\BibitemShut
  {NoStop}%
\bibitem [{\citenamefont {Mirzamohammadi}\ \emph {et~al.}(2025)\citenamefont
  {Mirzamohammadi}, \citenamefont {Pesquera}, \citenamefont {De~Luca},
  \citenamefont {Cordero-Edwards},\ and\ \citenamefont
  {Catalan}}]{Mirzamohammadi_apl25}%
  \BibitemOpen
  \bibfield  {author} {\bibinfo {author} {\bibfnamefont {N.}~\bibnamefont
  {Mirzamohammadi}}, \bibinfo {author} {\bibfnamefont {D.}~\bibnamefont
  {Pesquera}}, \bibinfo {author} {\bibfnamefont {G.}~\bibnamefont {De~Luca}},
  \bibinfo {author} {\bibfnamefont {K.}~\bibnamefont {Cordero-Edwards}},\ and\
  \bibinfo {author} {\bibfnamefont {G.}~\bibnamefont {Catalan}},\ }\bibfield
  {title} {\bibinfo {title} {Molecular electrodes enable faster switching in
  ferroelectric thin films},\ }\href {https://doi.org/10.1063/5.0257017}
  {\bibfield  {journal} {\bibinfo  {journal} {Applied Physics Letters}\
  }\textbf {\bibinfo {volume} {127}},\ \bibinfo {pages} {012904} (\bibinfo
  {year} {2025})}\BibitemShut {NoStop}%
\bibitem [{\citenamefont {Molina-Ruiz}\ \emph {et~al.}(2011)\citenamefont
  {Molina-Ruiz}, \citenamefont {Lopeand\'{\i}a}, \citenamefont {Pi},
  \citenamefont {Givord}, \citenamefont {Bourgeois},\ and\ \citenamefont
  {Rodr\'{\i}guez-Viejo}}]{Javier_prb11}%
  \BibitemOpen
  \bibfield  {author} {\bibinfo {author} {\bibfnamefont {M.}~\bibnamefont
  {Molina-Ruiz}}, \bibinfo {author} {\bibfnamefont {A.~F.}\ \bibnamefont
  {Lopeand\'{\i}a}}, \bibinfo {author} {\bibfnamefont {F.}~\bibnamefont {Pi}},
  \bibinfo {author} {\bibfnamefont {D.}~\bibnamefont {Givord}}, \bibinfo
  {author} {\bibfnamefont {O.}~\bibnamefont {Bourgeois}},\ and\ \bibinfo
  {author} {\bibfnamefont {J.}~\bibnamefont {Rodr\'{\i}guez-Viejo}},\
  }\bibfield  {title} {\bibinfo {title} {Evidence of finite-size effect on the
  n\'eel temperature in ultrathin layers of $\mathrm{CoO}$ nanograins},\ }\href
  {https://doi.org/10.1103/PhysRevB.83.140407} {\bibfield  {journal} {\bibinfo
  {journal} {Phys. Rev. B}\ }\textbf {\bibinfo {volume} {83}},\ \bibinfo
  {pages} {140407} (\bibinfo {year} {2011})}\BibitemShut {NoStop}%
\end{thebibliography}%

\section*{End Notes}

\subsection{Parameters}

We adopt here the standard material parameters for BaTiO$_3$~\cite{Book_Rabe07}
$a_1$=$3.34(T-381\,\mathrm{K})\times 10^5$\,C$^{-2}$m$^2$N,
$a_{11}$=$4.69(T-436\,\mathrm{K})\times 10^6$\,C$^{-4}$m$^6$N,
$a_{12}$=$3.2\times 10^8$\,C$^{-4}$m$^6$N,
%
%
$a_{111}$=$-(5.5(T-393\,\mathrm{K})+276)\times 10^7$\,C$^{-6}$m$^{10}$N,
$a_{112}$=$4.47\times 10^9$\,C$^{-6}$m$^{10}$N,
$a_{123}$=$4.92\times 10^9$\,C$^{-6}$m$^{10}$N,
$Q_{1111}$=$0.11$\,C$^{-2}$m$^4$,
$Q_{1122}$=$-0.045$\,C$^{-2}$m$^4$,
$Q_{1212}$=$0.015$\,C$^{-2}$m$^4$,
$C_{1111}$=$178\times 10^9$\,m$^{-2}$N,
$C_{1122}$=$96.4\times 10^9$\,m$^{-2}$N,
$C_{1212}$=$122\times 10^9$\,m$^{-2}$N.

\subsection{Polarization contribution}

The polarization renormalization is obtained by substituting the soft-domain profile, Eq.~(\ref{SoftDomain}), into Eq.~(\ref{FP}) and averaging the resulting polynomial terms over one domain period using the trigonometric identities for averaging:
\begin{equation}
\langle \sin^{2n}x\,\cos^{2m}x\rangle
=
\frac{(2n-1)!!(2m-1)!!}{(2n+2m)!!}.
\label{avg_rule}
\end{equation}
This gives the effective coefficients $\bar A_{\mathrm P}$, $\bar B_{\mathrm P}$, and $\bar C$, given in the main text by Eqs.~(\ref{AP}), (\ref{BP}), and (\ref{C}), respectively.

\subsection{Elastic contribution}

In the multidomain state, the spatially varying polarization profile of Eq.~(\ref{SoftDomain}) generates nonuniform internal stresses described by the dimensionless electrostrictive source tensor
\begin{equation}
\Pi_{ij}=p_i p_j,
\label{Pi}
\end{equation}
where the dimensionless polarization profile $\mathbf p=(p_1,p_2,p_3)=\mathbf P/P$ has components
$p_1=\gamma \sin\!\left( \kappa_x x \right)\sin\!\left(\kappa_z z\right)$,
$p_2=0$, and
$p_3=\cos\!\left( \kappa_x x \right)\cos\!\left(\kappa_z z\right)$.
The corresponding stress field must satisfy the mechanical equilibrium condition
\begin{equation}
\partial_j \sigma_{ij}=0,
\label{sigma-constrain}
\end{equation}
which makes the elastic problem nonlocal. As a result, the elastic energy cannot be obtained by direct minimization of $F_{\mathrm{elast}}$ with respect to the stress field, but requires an explicit solution of the elasticity equations under the mechanical equilibrium constraint. In addition, the elastic problem must be supplemented by a specification of the macroscopic (space-averaged) stress $\Bar{\sigma}_{ij}$, which is either fully relaxed or balances externally imposed forces, such as surface tension.

The problem is most conveniently treated in Fourier space, where linear elasticity decouples different stress modes and mechanical equilibrium is enforced independently for each wave vector. A further simplification follows from the periodicity of $\Pi_{ij}$, which inherits the spatial modulation of the polarization texture and can therefore be expanded into a finite set of plane-wave harmonics with five wave vectors,
\begin{gather}
\Pi_{ij}=\Pi^{(0)}_{ij}+\sum_{n=1}^4\Pi^{(n)}_{ij}\cos\bigl(\mathbf{k}^{(n)}\!\cdot\!\mathbf{r}\bigr),
\notag \\
\mathbf{k}^{(0)}=(0,0,0), \qquad
\mathbf{k}^{(1)}=(2\kappa_x,0,0), \qquad
\mathbf{k}^{(2)}=(0,0,2\kappa_z),
\notag \\
\mathbf{k}^{(3)}=(2\kappa_x,0,2\kappa_z), \qquad
\mathbf{k}^{(4)}=(2\kappa_x,0,-2\kappa_z).
\label{wavevectors}
\end{gather}
The corresponding coefficient matrices are
\begin{equation}
\Pi^{(n)}_{ij}=
\begin{pmatrix}
a_n & 0 & c_n\\
0 & 0 & 0\\
c_n & 0 & b_n
\end{pmatrix},
\qquad n=0,\dots,4,
\label{Pmatrix}
\end{equation}
with
$a_{0}=\frac{1}{4}\gamma^2$,
$a_{1,2}=-\frac{1}{4}\gamma^2$,
$a_{3,4}=\frac{1}{8}\gamma^2$,
$b_{0,1,2}=\frac{1}{4}$,
$b_{3,4}=\frac{1}{8}$,
$c_{0,1,2}=0$,
$c_{3}=-\frac{1}{8}\gamma$,
$c_{4}=\frac{1}{8}\gamma$.

Introducing
\begin{equation}
U_{ij}^{(n)}=Q_{ijkl}\Pi_{kl}^{(n)},
\label{Uij}
\end{equation}
the elastic energy, Eq.~(\ref{Felast}), can be written in Fourier representation as
\begin{gather}
F_{\mathrm{elast}}
=-\,\Bar{\sigma}_{ij}\,U_{ij}^{(0)}P^2
\notag \\
-\sum_{n=1}^4\left[
\frac{1}{2}\,\sigma_{ij}^{(n)}U_{ij}^{(n)}P^2
+
\frac{1}{4}\,S_{ijkl}\sigma_{ij}^{(n)}\sigma_{kl}^{(n)}
\right].
\label{FelastFourier}
\end{gather}
The zero-harmonic term yields the quadratic correction $\Bar A_{\mathrm{elast}}$, given in the main text by Eq.~(\ref{Ael}) and determined by the macroscopic stress, while the nonzero harmonics produce the quartic elastic self-energy correction $\Bar B_{\mathrm{elast}}$, as given by Eq.~(\ref{Bel}) in the main text. These two contributions are discussed separately below.

\subsection{Quadratic renormalization from surface tension}

We consider the case where the macroscopic stress is generated by in-plane stress  so that the in-plane components of the space-averaged stress are fixed as
\begin{equation}
\Bar{\sigma}_{xx}=\Bar{\sigma}_{yy}=\sigma_s,
\label{sigma_macro}
\end{equation}
The resulting quadratic elastic correction follows from the zero-harmonic part of Eq.~(\ref{FelastFourier}),
\begin{equation}
\Bar{A}_{\mathrm{elast}}=-\,\Bar{\sigma}_{ij}\,U_{ij}^{(0)},
\label{CoefElA}
\end{equation}
and yields the expression given in the main text, Eq.~(\ref{Ael}). 

\subsection{Quartic renormalization from elastic self-energy}

\paragraph*{Physical origin of the nonlocal quartic term.}
The quartic elastic correction originates from the self-energy of the nonzero stress harmonics in Eq.~(\ref{FelastFourier}). Physically, this contribution is nontrivial because the modulated electrostrictive sources cannot relax independently at each point in space: the induced stress field must satisfy mechanical equilibrium condition~(\ref{sigma-constrain})  and is therefore constrained globally. 

As shown below, the effective elastic correction to the quartic coefficient, $\Bar{B}_{\mathrm{elast}}$, given in the main text by Eq.~(\ref{Bel}), can be written as
\begin{equation}
\Bar{B}_{\mathrm{elast}}
=
\frac14\,\sum_{n=1}^4
{C}^{\mathbf{k}^{(n)}}_{ijml}\,
U_{ij}^{(n)}U_{ml}^{(n)},
\label{CoefBEl}
\end{equation}
The essential point is the evaluation of the kernels ${C}^{\mathbf{k}^{(n)}}_{ijml}$, which determine how the bare elastic response is modified by the mechanical-equilibrium constraint for each Fourier harmonic. This evaluation involves constrained tensorial algebra, as outlined below.

\paragraph*{Constrained minimization in Fourier space.}
The kernels ${C}^{\mathbf{k}^{(n)}}_{ijml}$ are obtained by minimizing the nonuniform part of the elastic energy, $F_{\mathrm{elast}}(\sigma_{ij}^{(n)})$, in Eq.~(\ref{FelastFourier}), subject to the mechanical-equilibrium constraint~(\ref{sigma-constrain}). In Fourier space, this constraint reads
\begin{equation}
k_i^{(n)}\sigma_{ij}^{(n)}=0,
\label{k-constrain}
\end{equation}
and is enforced by the Lagrange multipliers $\lambda_i^{(n)}$. The corresponding effective Lagrange functional then takes the form
\begin{equation}
\tilde F_{\mathrm{elast}}
=
F_{\mathrm{elast}}(\sigma_{ij}^{(n)})
+
\sum_{n=1}^{4}\lambda_i^{(n)} k_j^{(n)} \sigma_{ij}^{(n)}.
\label{Lagrange}
\end{equation}

Minimization with respect to $\sigma_{ij}^{(n)}$ gives
\begin{equation}
\sigma_{ij}^{(n)}
=
-\,C_{ijmn}U_{mn}^{(n)}P^2
-2\,C_{ijmn}\lambda_m^{(n)}k_n^{(n)},
\label{sigma_general}
\end{equation}
where $C_{ijmn}=(S^{-1})_{ijmn}$. Imposing Eq.~(\ref{k-constrain}) yields
\begin{gather}
\lambda_m^{(n)}
=
-\frac{1}{2}\left(\Gamma^{-1}\right)_{mr}^{(n)}
k_j^{(n)}C_{rjpn}U_{pn}^{(n)}P^2,
\label{lambda}
\\
\Gamma_{rm}^{(n)}=k_v^{(n)}C_{vrtm}k_t^{(n)}.
\label{gamma}
\end{gather}

\paragraph*{Constrained elastic kernel.}
Substituting this result back into Eq.~(\ref{sigma_general}), one obtains the projected stress form
\begin{equation}
\sigma_{ij}^{(n)}
=
-\,P^2\,C_{ijmn}^{\mathbf{k}^{(n)}}U_{mn}^{(n)},
\label{sigma_projected}
\end{equation}
with
\begin{equation}
C_{ijmn}^{\mathbf{k}^{(n)}}
=
C_{ijmn}
-
C_{ijrs}k_s^{(n)}(\Gamma^{-1})^{(n)}_{rt}k_u^{(n)}C_{tumn},
\label{Ck_def}
\end{equation}
where $\Gamma^{(n)}_{rt}$  is given by Eq.~(\ref{gamma}).
This is the effective kernel that governs the quartic elastic correction in Eq.~(\ref{CoefBEl}).
Thus, we obtain an analytical representation of the quartic elastic correction $\Bar B_{\mathrm{elast}}$ in terms of the system parameters. Its fully explicit closed form is, however, unwieldy in practice, since the calculation requires repeated contractions and inversions of multidimensional tensorial matrices for each Fourier harmonic. We therefore evaluated $\Bar B_{\mathrm{elast}}$ using a dedicated MATLAB implementation, which is provided in ~\cite{LukyanchukBoron2026Zenodo}.

\newpage
\onecolumngrid

\begin{center}
\underline{\textbf{\large Supplemental Material}}
\end{center}

In the Supplemental Material, we discuss the fast scanning calorimeter for free-standing membranes, covering sample preparation, membrane transfer, measurement techniques, the thermal link connecting the sample to the calorimeter, and finally the data analysis for extracting latent heat and adiabatic heat capacity in detail.

\section{Nano-calorimetry of free-standing membranes}

The membrane-based fast scanning calorimeters allow measurement of the nearly adiabatic (or semi-adiabatic) heat capacity of nanoscale samples, ranging from organic semiconductors, polymers, and metal alloys to metallic glasses and various other materials \cite{Allen_ThermoActa04, Rodríguez-Viejo_Book16}. The method has proved to be very efficient for understanding size effects in metallic or magnetic transitions  \cite{Javier_prb11}, as well as the kinetics of polymers and organic materials \cite{Vila-Costa_prl20}. However, in most reported studies, heat capacity measurements have been limited to polycrystalline or amorphous structures that are directly grown on the calorimetric cell \cite{Javier_prb11}, and the requirement for a specific substrate for the epitaxial growth of monocrystals has restricted its use until now.

The recent advancements in free-standing complex oxide membranes, following the graphene revolution, provide enormous opportunities to study perfectly crystalline nanomaterials from different perspectives \cite{Lu_NatMat16, David_AdvMat20, Pesquera_JPCM22}. While surface and electrical studies on the nanoscale are common, direct thermodynamic measurements that scale with the sample volume are rare. Here, we merge the chemical lift-off technique with nanocalorimetry to study the ferroelectric phase transition of free-standing BaTiO$_3$ membranes.\\

\begin{figure*}[h!]
\centering
\includegraphics[width=0.95\textwidth]{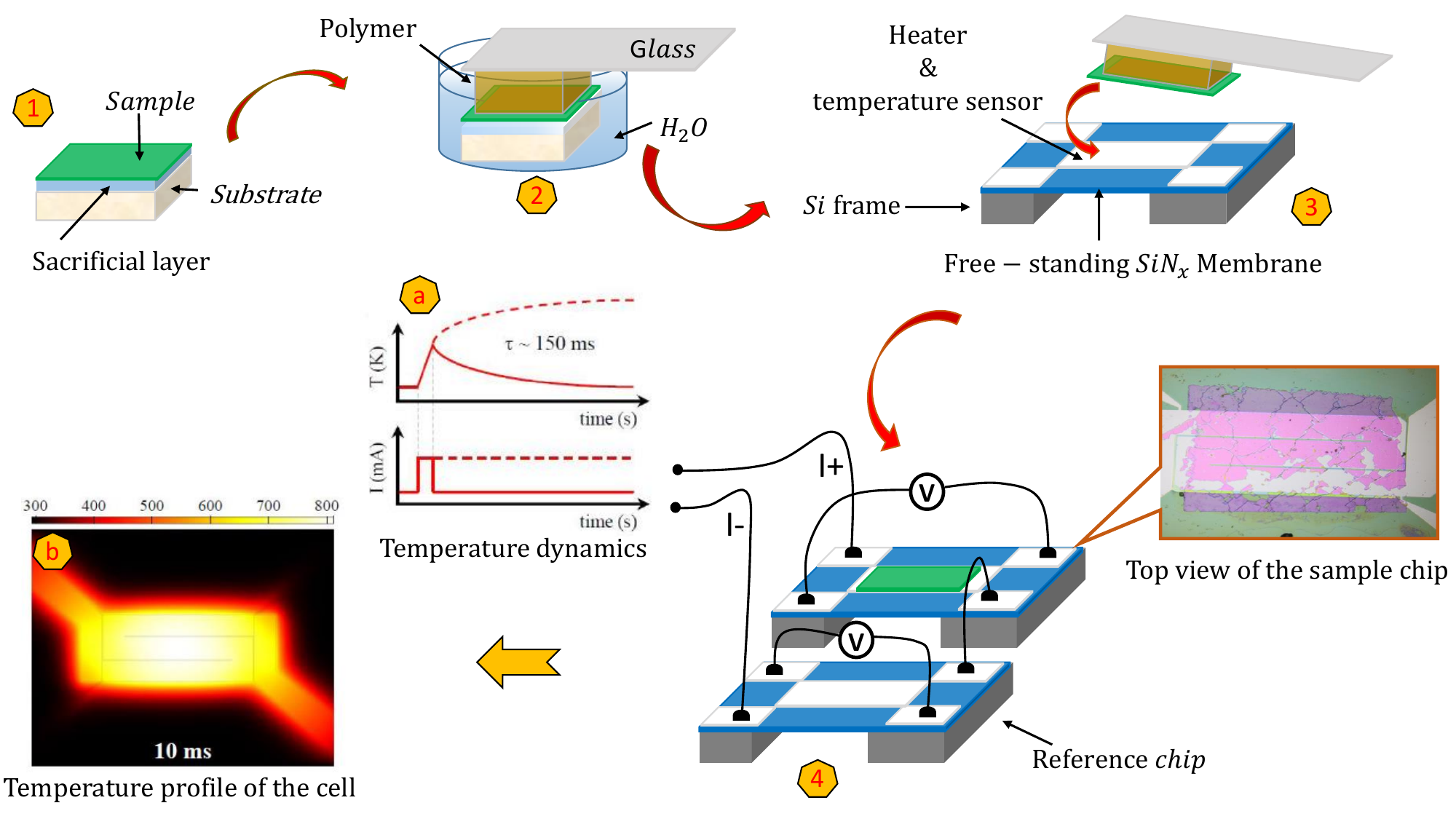}
\caption[0.5\textwidth]{Schematic diagram of nanocalorimetry for free-standing membranes. (1,2) Release the sample from the substrate by dissolving the sacrificial layer in water. (3) Stamping the sample onto the calorimeter's sensing area using a organic semiconductor (TPD) coated polymer (PDMS). (4) Schematic of the heat capacity measurement in differential mode by applying a short current pulse and recording the voltage drop across the calorimetric cell. (a) Temperature dynamic in the nanocalorimetry due to the current pulse. (b) COMSOL Multiphysics simulation of the temperature profile in a calorimetric cell during a heating scan at $8\times10^4$ K/s.}
\label{fig:Nano_cal}
\end{figure*}
 
\textbf{Sample preparation:} The sample was grown using the pulsed laser deposition (PLD) technique on a (100) GdScO$_3$ substrate placed on a heating plate at 750$^{\circ}$C. The distance between the stoichiometric ceramic target and the substrate was 55 mm. A KrF excimer laser ($\lambda = 248$ nm) was focused on the rotating target (15 rpm) using a 300 mm single convex lens, producing a fluence of 1.825 J/cm$^2$ at a repetition rate of 2 Hz. During growth, the oxygen pressure was maintained at 100 mTorr. After growth, the oxygen pressure was immediately increased to 10 Torr, and the temperature was slowly decreased at a cooling rate of 5 K/min.

The membranes were exfoliated from the original substrate by breaking the chemical bonds of a sacrificial buffer layer grown between the substrate and the oxide film. Sr$_3$Al$_2$O$_6$ was chosen as the buffer layer because its lattice parameters are similar to those of the BaTiO$_3$ film, and it easily dissolves in water, making it an ideal sacrificial layer for BaTiO$_3$ membrane fabrication. A few nanometers (5-7 nm) of Sr$_3$Al$_2$O$_6$ were epitaxially grown between the GdScO$_3$ substrate and the BaTiO$_3$ film [Fig. \ref{fig:Nano_cal}(1)].

We attached the film–substrate combination to a polymer (PDMS), gripped by a glass slide, and immersed it in water [Fig. \ref{fig:Nano_cal}(2)]. A 100 nm organic semiconductor TPD [N,N-Diphenyl-N,N'-bis(methylphenyl)-1,1'-biphenyl-4,4'-diamine] was vapor-deposited on top of a flat poly-dimethylsiloxane (PDMS) layer to increase the efficiency of the transfer process. Due to the small size of the organic molecules, any residual TPD left after the transfer was easy to remove, ensuring contamination-free, strain-free, high-quality BaTiO$_3$ membranes. The sacrificial Sr$_3$Al$_2$O$_6$ buffer layer dissolved in water, resulting in a substrate-free BaTiO$_3$ film attached to the polymer.

The BaTiO$_3$ film was then rubber-stamped onto the sensing area of the calorimeter by heating at 80$^{\circ}$C for 10 min [Fig. \ref{fig:Nano_cal}(3)]. Since the glass transition temperature of TPD is about 60$^{\circ}$C, the polymer becomes less adhesive to the film, making the lift-off process more efficient than usual. Because of the small size and non-repetitive structure of TPD, any residue on the film could be easily removed by evaporating at 150$^{\circ}$C and rinsing in acetone. Finally, we strengthened the attachment of the film to the calorimeter by heating at 120–140$^{\circ}$C for 15–20 min. A representative image of BaTiO$_3$ membranes after transfer onto the nano-calorimetric chip is shown in Fig. \ref{fig:sample}(a). The thermal link between the sample and the calorimeter was sufficiently strong to perform fast scanning measurements. A detailed description of the thermal link is provided in the next section of this supplementary document.\\ 

\textbf{Nano-calorimetry:} The nano-calorimetry technique, developed a few years ago and described in detail in previous studies \cite{Allen_ThermoActa04, Rodríguez-Viejo_Book16}. Typically, a thin (150 nm) platinum layer deposited on top of a 450 nm free-standing SiN$_x$ membrane, which serves as a heater. The free-standing Si$_3$N$_4$ membrane is supported by a Si frame to reduce thermal conduction. Localized Joule heating is achieved by injecting short current pulses (20–60 mA for 2–20 ms) under high vacuum ($\approx$ 10$^{-6}$ mbar), rapidly increasing the membrane temperature at rates of 10$^3$–10$^5$ K/s. Simultaneously, the rise in resistance of the metal film due to heating allows the membrane temperature to be measured by monitoring the voltage across the sensing area, with proper pre-calibration. The serpentine geometry of the heater ensures a uniform temperature profile at the center of the calorimetric cell (sensing area), even at very high heating rates [Fig. \ref{fig:Nano_cal}(b)].

Generally, nano-calorimetry is performed in differential mode to increase the sensitivity of the measurements when the mass of the sample is sufficiently small ($<$ 100 ng) [Fig. \ref{fig:Nano_cal}(4)]. In this case, a reference calorimetric chip nearly identical to the sample chip (after transferring the sample) is connected in series, allowing the heat capacity of the sample to be obtained directly with a minimal contribution from the calorimeter addenda, and therefore dramatically increasing the accuracy. When the sample mass is large and differential mode operation is not feasible, measurements are carried out in single-mode configuration, wherein a single calorimeter is used and the heat capacity of the empty device is subtracted afterward.

One of the main advantages of calorimetry in the context of this work comes from the fact that calorimetric measurements depend on the volume, or more precisely, the mass of the sample—the higher the mass, the higher the heat capacity. Therefore, the calorimetric measurements are not significantly influenced by the surface of the membrane, as is the case in microscopic and electrical characterization techniques. This is especially important considering that the morphology of the transferred oxide membrane can be significantly affected by the transfer process.

Furthermore, nanocalorimetry, owing to the high heating rates imposed and the thin membrane serving as the calorimetric cell, effectively minimizes heat losses by conduction. In addition, because the measurements are performed under high-vacuum conditions, quasi-adiabatic heat capacity measurements can be obtained. Although completely adiabatic measurements are virtually impossible—since minimal heat losses to the surroundings are inevitable—a reliable estimation of the true adiabatic heat capacity can be achieved by extrapolating the measured heat capacity at different heating rates to the limit of infinite rate.\\

\begin{figure}
\centering
\includegraphics[width=0.7\textwidth]{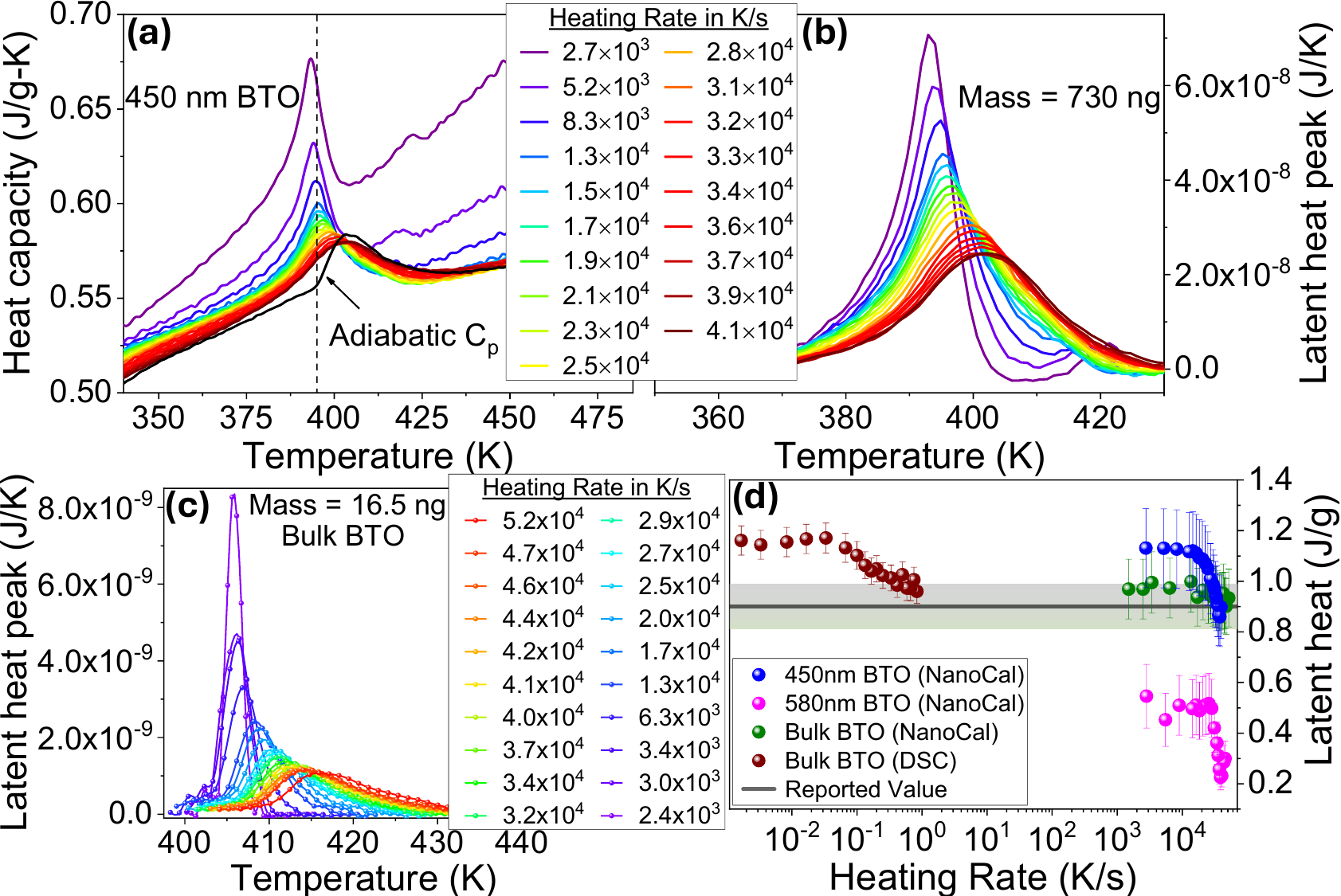}
\caption[0.5\textwidth]{(a) Heat capacity of the 450 nm-thick membrane measured at different heating rates. The latent heat peaks of the ferroelectric transition for the 450 nm BaTiO$_3$ membrane (b) and bulk sample (c), extracted by subtracting the background heat capacity. (d) The latent heat values for membranes and bulk samples are presented in comparison with the range of reported values for the BaTiO$_3$ ceramic. Bulk measurements at lower heating rates were performed using a power-compensated commercial differential scanning calorimeter (DSC).}
\label{latentheat}
\end{figure}

\textbf{Adiabatic heat capacity:} The genuine adiabatic heat capacity is extracted by accounting for heat losses due to the finite heating rate. This is calculated by linearly extrapolating rate-dependent measurements at each temperature over the full measurement range [Fig. \ref{latentheat}(a)]. The linear extrapolation is valid provided there is a strong thermal link between the sample and the calorimeter. The thermal resistance can be estimated through simulations using COMSOL Multiphysics software. Also, we experimentally observed a strong thermal link in our system, which is presented at the end of the section. A small thermal lag between sample and the calorimeter is responsible for shifts in the onset temperature at higher heating rates.

While this extrapolation method can be employed to determine the real onset temperature of the transition, as summarized in Table \ref{table}, it becomes unreliable within the phase transition region, since the heat capacity at a first-order phase transition is undefined. Consequently, the adiabatic latent heat cannot be extracted using this extrapolation technique, as the peaks in the extrapolated heat capacity [black line in Fig. \ref{latentheat}(a)] are not physically meaningful.\\

\begin{figure*}[h!]
\centering
\includegraphics[width=0.7\textwidth]{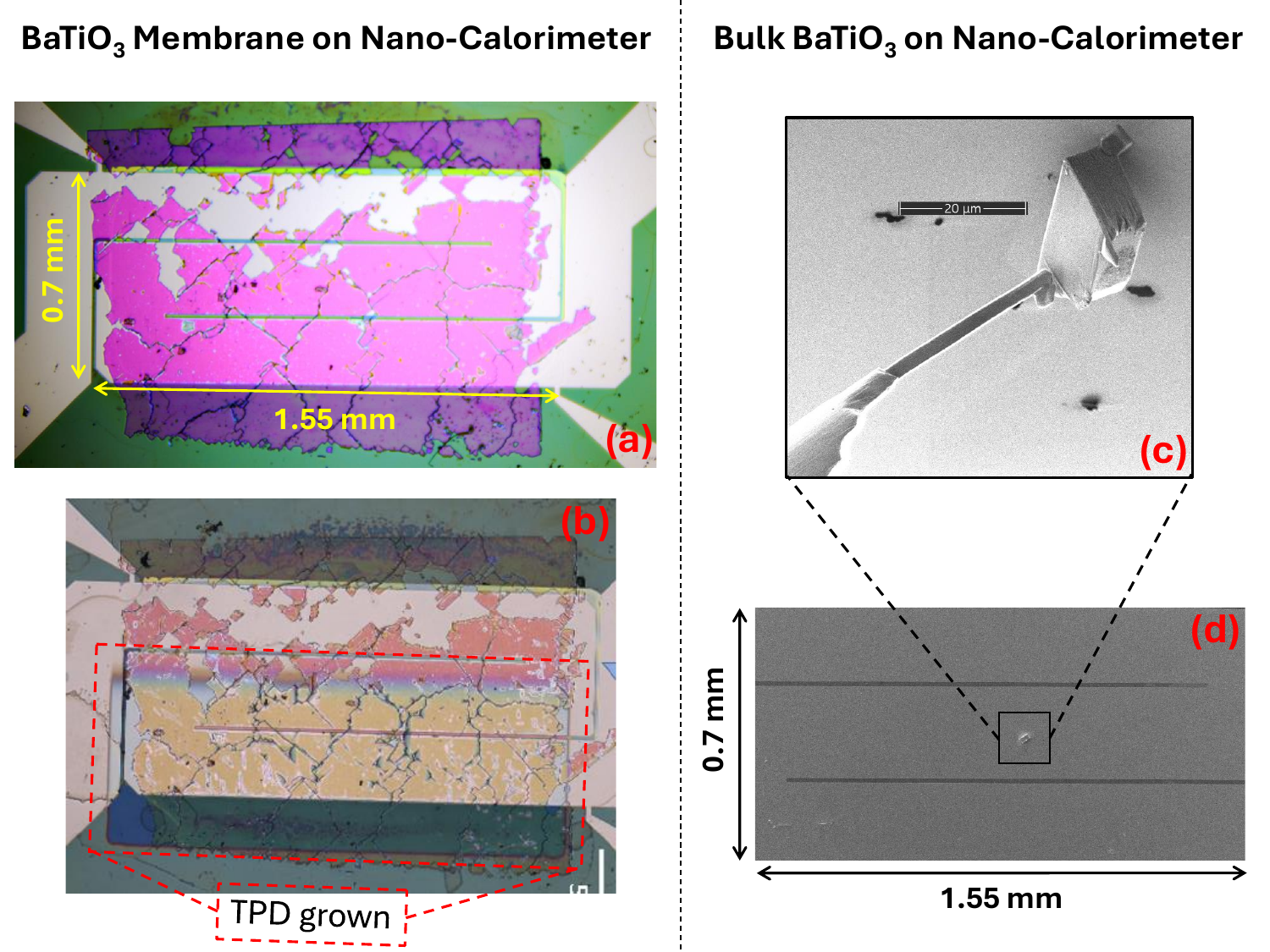}
\caption[0.5\textwidth]{(a) A representative image of BaTiO$_3$ membranes after transfer onto the nano-calorimetric chip. (b) Image of a calorimetric chip where a 90 nm organic semiconductor, TPD, has been grown on top of the 140 nm BaTiO$_3$ membrane. A proper mask was used to ensure that TPD grew only on top of the membrane and not on the open sensing areas of the calorimeter. This image indicates that the temperature experienced by TPD corresponds to the temperature at the top surface of the BaTiO$_3$ membrane. (c) Image of a FIB-cut bulk BaTiO$_3$ sample (volume $\approx 19\times14\times10\ \mu m^3$ and mass $\approx 16\ ng$) during the transfer process on to the nano-calorimeter. (d) The final image shows the sensing area of the calorimeter after transferring the FIB-cut bulk sample.}
\label{fig:sample}
\end{figure*}

\textbf{Latent heat extraction:} The latent heat peaks for different measurements were obtained by subtracting the background heat capacity, assuming the same temperature trend as observed outside the transition regions. The latent heat peaks at different heating rates are shown in Fig. \ref{latentheat}(b, c) for the 450 nm BTO membrane and the bulk sample. The integrated area of the heat capacity peak corresponds to the total heat released during the transition. For the 450 nm membrane, the latent heat is in good agreement with the bulk value over a wide range of heating rates [Fig. \ref{latentheat}(d)]. Bulk latent heat measurements at lower heating rates were also carried out using a conventional Differential Scanning Calorimetry (DSC) setup (DSC250 from TA Instruments). The heat losses in DSC are significant, and large sample masses are required; thus, for free-standing membranes, heat flow measurements at low heating rates are not feasible. The lower heating rate data are presented only for a single crystal with a relatively large mass (24 mg). Previously reported values at lower heating rates (1–30 K/min) \cite{Mathur_AdvMat13} fall within the shaded region of Fig. \ref{latentheat}(d). The semi-adiabatic latent heat of bulk BaTiO$_3$ single crystals was measured using nano-calorimetry in combination with a focused ion beam (FIB). A small piece of the sample was cut and transferred onto the nano-calorimetric chip [Fig. \ref{fig:sample}(c, d)]. During the transfer process, a small amount of tungsten alloy was used to attach the sample to the calorimeter. Subsequently, the heat capacity of the sample was measured using differential-mode nano-calorimetry. Due to an additional, unknown heat capacity contribution from the tungsten alloy, determining the heat capacity of the bulk BaTiO$_3$ sample with this technique was not feasible. However, the latent heat peak [Fig. \ref{latentheat}(c)] originates solely from the BaTiO$_3$ sample. The obtained semi-adiabatic latent heat value is consistent with previously reported values at lower heating rates [Fig. \ref{latentheat}(d)].

It is important to note that the latent heat value for the 570 nm membrane is about 40$\%$ of the bulk value [Fig. \ref{latentheat}(d)], while for the 50 nm, 140 nm and 300 nm membranes (shown in Fig 1 in the main text of the article) the latent heat peaks are not visible, and the values are therefore below 2$\%$, which is the measurement accuracy of our setup. The reduction in latent heat compared to the bulk value observed for the 570 nm membrane can be attributed to two factors. First, transferring the thicker membrane onto the calorimetric chip became increasingly difficult with thickness, and several repeated transfer attempts were required for the 570 nm sample. As a result, the sample surface was not as clean as that of the 450 nm or thinner membranes. The presence of additional, uncharacterized material can contribute to the measured heat capacity but not to the latent heat, leading to an apparent reduction. Second, we observed that the domain configuration of the membrane changed from monodomain to polydomain due to the repeated transfer attempts, which influence the measured transition enthalpy, consistent with the central argument of this article. The qualitative results suggest that for thicker membranes (thickness $d > 400$ nm), the transition is first-order, whereas for thinner films the transition is second-order. The specific heat jump anomaly at the ferroelectric transition of BTO crystals is very small. It can be resolved in sufficiently thin samples by differential measurements \cite{Javier_prb11}, but its detection in single-sample thermodynamic measurements, such as those used here, is much more demanding because of limited sensitivity. Nonetheless, the two different slopes of the specific heat below and above the transition region clearly indicate two distinct phases of the material [see Fig. 1(a, c) (main text)].\\

\begin{table}[h]
\begin{tabular}{|l|l|l|}
\hline
Samples & Measurement Technique & Onset Temperature\\
&&\\
\hline
450 nm BTO Membranes & Nanocalorimetry & 394 $\pm$ 2 (K)\\
&&\\
\hline
580 nm BTO Membranes & Nanocalorimetry & 396 $\pm$ 2 (K)\\
&&\\
\hline
Bulk BTO & Nanocalorimetry & 405 $\pm$ 2 (K)\\
& Conventional DSC & 402 $\pm$ 1 (K)\\
\hline
\end{tabular}
\caption{Onset temperature obtained from nanocalorimetry measurements during heating for different BTO samples. The onset temperature for the bulk sample was also measured at a very slow scanning rate using a conventional DSC setup.}
\label{table}
\end{table}

\begin{figure}
\centering
\includegraphics[width=0.47\textwidth]{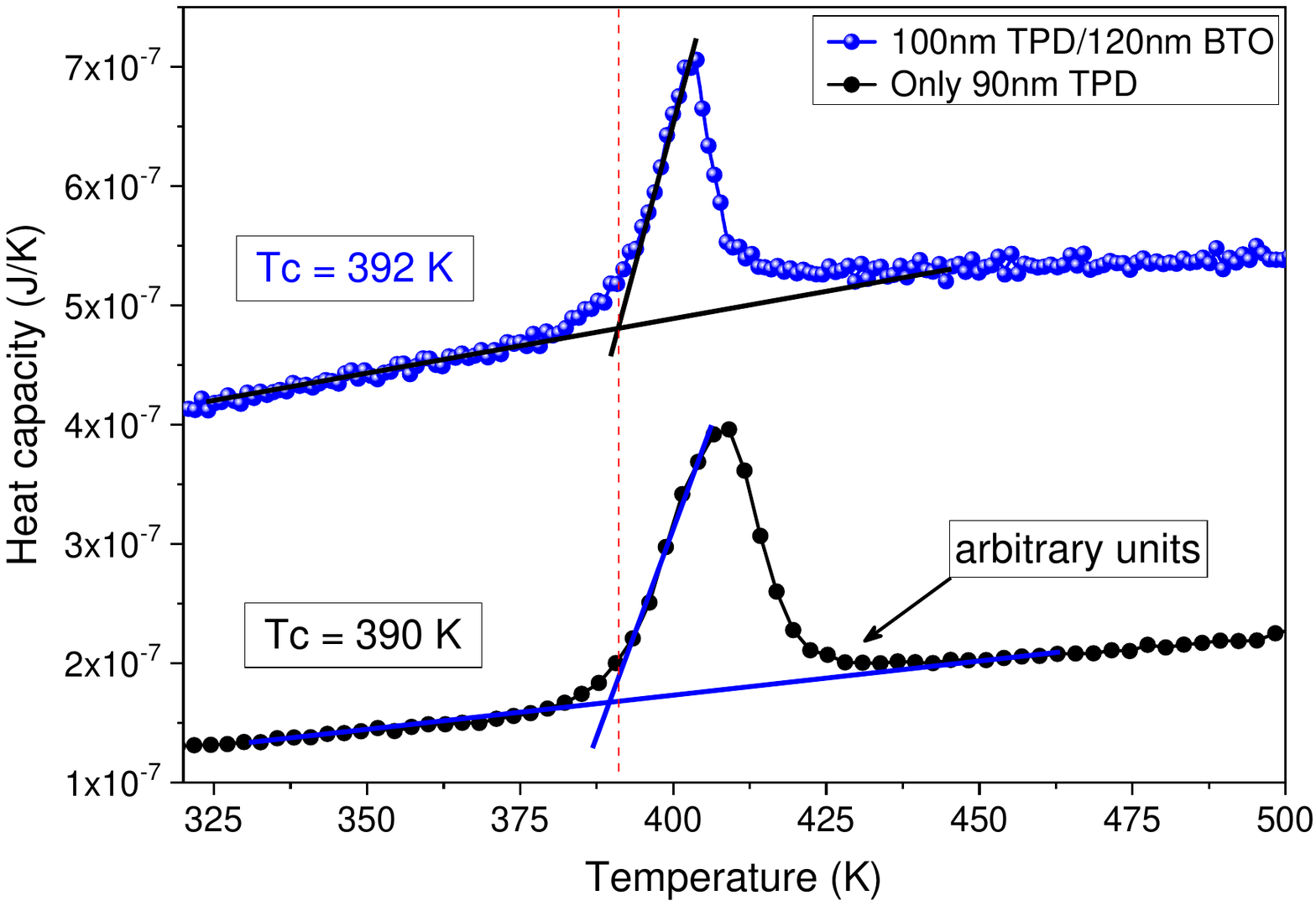}
\caption[0.5\textwidth]{Nano-calorimetry measurements performed on 100 nm TPD deposited directly on the chip (black) and 90 nm TPD grown on top of a 120 nm BaTiO$_3$ sample attached to the chip. The melting of the organic semiconductor TPD produces a peak in the heat capacity graph. The onset of the peak (T$_c$) can be considered as the melting point of the corresponding TPD.}
\label{fig:TPD_BTO}
\end{figure}

\textbf{Thermal link:} The thermal resistance between the sample and the calorimeter plays a crucial role in fast-scanning calorimetry. Even a small thermal resistance can create a significant temperature delay between the sample and the thermometer. To extract the true sample (transition) temperature, it is necessary to account for this thermal lag.

To measure the thermal lag between the BaTiO$_3$ membrane and the calorimeter, we performed nano-calorimetry measurements on a 90 nm TPD layer grown on top of the BaTiO$_3$ sample placed on the sensing area of the calorimeter [Fig. \ref{fig:sample} (b)], and on a 100 nm TPD sample grown directly on the calorimetric chip. The temperature increase causes the organic semiconductor TPD to melt, and the melting temperature depends on thickness, heating rate, growth temperature, and growth environment \cite{Vila-Costa_prl20}. To compare the melting points, nano-calorimetry of TPD in both cases was carried out in an in-situ platform at similar heating rates and under identical growth conditions.

Figure \ref{fig:TPD_BTO} shows that there is no significant difference in the melting temperatures of TPD due to the presence of the BaTiO$_3$ membrane between the calorimeter and the TPD sample. These results suggest that, although the measurements are performed at very high rates, the thermal lag between the sample and the calorimeter is negligible, allowing reliable nano-calorimetry measurements on such membranes. The high thermal linkage is likely due to the strong van der Waals forces between the platinum heater and the BaTiO$_3$ sample.

\end{document}